\documentclass[lettersize,journal]{IEEEtran}
\usepackage{amsmath,amsfonts}
\usepackage{algorithmic}
\usepackage{algorithm}
\usepackage{array}
\usepackage[caption=false,labelfont=sf,textfont=sf]{subfig}
\usepackage{textcomp}
\usepackage{stfloats}
\usepackage{url}
\usepackage{verbatim}
\usepackage{graphicx}
\usepackage{cite}
\usepackage{multirow}
\usepackage{booktabs}
\usepackage{pgf}
\usepackage{tcolorbox}
\tcbuselibrary{breakable}  
\usepackage{makecell}
\usepackage{hyperref}
\usepackage{cleveref}
\crefname{section}{§}{§§}

\usepackage{xcolor}
\definecolor{highlightred}{HTML}{e41a1c}
\definecolor{highlightorange}{HTML}{ff7f00} 
\definecolor{highlightyellow}{HTML}{ffff33} 
\definecolor{highlightgreen}{HTML}{4daf4a}
\definecolor{highlightper}{HTML}{984ea3}
\definecolor{highlightblue}{HTML}{377eb8}
\definecolor{highlightbrown}{HTML}{bf5b17}

\usepackage{colortbl} 

\newtcolorbox{listbox}[1][]{
    nobeforeafter,
    colframe=white,  
    colback=green!10,  
    boxsep=0pt,  
    left=0pt, right=0pt, top=0pt, bottom=0pt,
    arc=0pt,  
    boxrule=0pt,  
    width=\linewidth,  
    enlarge left by=-\fboxsep,
    enlarge right by=-\fboxsep,
    #1  
}

\usepackage{tikz}

\begin{document}

\title{On the Reliability of Biometric Datasets: \\How Much Test Data Ensures Reliability?}

\author{Matin Fallahi\textsuperscript{1}, Ragini Ramesh\textsuperscript{2}, Pankaja Priya Ramasamy\textsuperscript{2}, Patricia Arias Cabarcos\textsuperscript{1,2}, \\ Thorsten Strufe\textsuperscript{1}, Philipp Terhörst\textsuperscript{2} \\
\textsuperscript{1}KASTEL Security Research Labs - KIT, Karlsruhe, Germany \\
\textsuperscript{2}University of Paderborn, Paderborn, Germany
}

\markboth{Journal of \LaTeX\ Class Files,~Vol.~14, No.~8, August~2021}%
{Shell \MakeLowercase{\textit{et al.}}: A Sample Article Using IEEEtran.cls for IEEE Journals}

\maketitle

\begin{abstract}
Biometric authentication is increasingly popular for its convenience and accuracy. However, while recent advancements focus on reducing errors and expanding modalities, the reliability of reported performance metrics often remains overlooked. Understanding reliability is critical, as it communicates how accurately reported error rates represent a system's actual performance, considering the uncertainty in error-rate estimates from test data. Currently, there is no widely accepted standard for reporting these uncertainties and indeed biometric studies rarely provide reliability estimates, limiting comparability and interpretation. To address this gap, we introduce BioQuake—a measure to estimate uncertainty in biometric verification systems—and empirically validate it on four systems and three datasets. Based on BioQuake, we provide simple guidelines for estimating performance uncertainty and facilitating reliable reporting. Additionally, we apply BioQuake to analyze biometric recognition performance on 62 biometric datasets used in research across eight modalities: face, fingerprint, gait, iris, keystroke, eye movement, Electroencephalogram (EEG), and Electrocardiogram (ECG). Our analysis shows that reported state-of-the-art performance often deviates significantly from actual error rates, potentially leading to inaccurate conclusions. To support researchers and foster the development of more reliable biometric systems and datasets, we release BioQuake as an easy-to-use web tool for reliability calculations.
\end{abstract}


\section{Introduction}
\label{sec:Introduction}
Biometric verification allows for the automatic verification of a person's identity based on their biological (e.g., face, fingerprint) or behavioral characteristics (e.g., keystroke dynamics) \cite{ISO2382-37}. This type of user authentication plays a key role in providing secure access to our devices, as well as to the many physical and digital environments we navigate daily \cite{ometov2018multi}. Indeed, rapid advances in the accuracy and processing speed of biometric systems have facilitated their widespread adoption across a range of high-security applications, such as border control, forensics, and financial transactions, as well as their integration into smartphones and other consumer devices. This trend is expected to grow building on the increased research in the area. State-of-the-art literature is intensive on exploring new, potentially more usable and robust, biometric modalities {\cite{sluganovic2018analysis, eberz201928, melzi2023ecg,arias2023performance, maiorana2021learning, chu2019ecg}} that can be incorporated in current systems or tailored to emerging technology, such as extended reality  \cite{meng2014surveying, stephenson2022sok}. Another branch of research focuses on improving the performance of well-established biometrics in different environmental conditions {\cite{liu2024latent, terhorst2023qmagface, meng2021magface, wang2019racial, zheng2017cross}.}
 
 Whether developing new biometric systems or improving existing techniques, there is a critical need for comprehensive and reliable testing. However, most of the proposed solutions are tested on small datasets and/or provide insufficient information about performance and reliability of the reported results \cite{sugrim2019robust}. These issues hinder rapid advance in biometric research, as it is not clear how to compare the performance of different systems to verify if there is a true improvement. To better understand the problem, let’s examine the common reporting practice. Since biometric matching is probabilistic, the ISO standard on \textit{``Biometric performance testing and reporting''} \cite{isoiec19795-1:2021}  recommends using error rates: (1) for incorrect matches where impostors are identified as genuine users (False Match Rate, FMR), and (2) for incorrect rejections where genuine users are identified as impostors (False Non-Match Rate, FNMR). {Although insightful, these metrics do not account for how closely the observed error rates reflect true error rates, due to testing data limitations (e.g., insufficient size or diversity). For example, take two systems that report the same False Match Rate (FMR) of 0.1\%. If one system consistently performs near this rate while the other fluctuates significantly, the latter could pose major security risks. This is because such variability increases the likelihood of unexpected impostor acceptances.}

Uncertainty estimations are needed to quantify how much the observed error rate might deviate from the true one, and should be reported for a more accurate representation of the results  \cite{jolliffe2007uncertainty,gilleland2010confidence}. Several studies \cite{dass2006validating,veres2006enough,li2012test, shen1997evaluation, bolle2004error} provided approaches to quantify the relationship between uncertainty and biometric verification performance (\cref{sec:RelatedWork}). However, these works have two main limitations. First, they rely on strong assumptions about the data, such as that the samples follow a normal distribution and are uniform across subjects, which may not hold in real-world scenarios. Second, these methods are challenging to implement in practice due to their complexity and the need for specialized statistical expertise, making them less accessible for practical application. Indeed, despite these uncertainty quantification approaches exist, biometric research very rarely reports uncertainties. In this paper, we bridge these gaps through the following contributions:

\begin{enumerate}
    \item We formalize \textbf{BioQuake \footnote{The name \textit{BioQuake} was chosen to reflect the potential `shaking' or variability (\textit{Quake}) in the observed performance metrics of biometric systems (\textit{Bio}) due to uncertainty.},  a new uncertainty metric for biometric systems}. BioQuake (\cref{sec:MethodologyRU}) describes how much the true error rate might deviate from the observed error (FMR, FNMR), given a specified confidence level and the number of genuine/impostor comparisons. To validate the metric, we conduct experiments on three biometric datasets for face recognition~\cite{phillips1998feret,huang2008labeled,eidinger2014age}, each containing over ten thousand images. These datasets are frequently used in state-of-the-art research and were captured under varying conditions. We apply four face recognition systems to these datasets and demonstrate that BioQuake is an accurate estimator of empirical uncertainty by measuring the correlation between empirical measurements and theoretical uncertainty  (\cref{sec:ExperimentalSetup_ee}, \cref{sec:resultsCorrectness}). BioQuake treats sample-pairs as independent and identically distributed and models uncertainty using a binomial distribution of these pairs, which is less restrictive than existing metrics that assume a normal distribution of the individual samples. While its accuracy may be influenced by factors such as a small number of subjects relative to the total samples, or an uneven distribution of samples among subjects, BioQuake provides a reliable lower bound on uncertainty. We further discuss these considerations and its applicability in real-world scenarios in \cref{sec:limitations}.
    
    \item We introduce \textbf{easy-to-apply uncertainty estimation rules and a user-friendly calculation tool}. Based on the theoretical definition of BioQuake, we derive three heuristics that are helpful for conveniently estimating the uncertainty of a biometric system's performance or the number of comparisons needed to achieve a specific reliability (\cref{sec:MethodologyRules}). These rules are derived for a confidence level of 95\% and ensure that the true error only deviates by either 1\%, 6\%, or 10\%. Additionally, we implemetned and released an online tool\footnote{\url{http://i63fallahi.ps.kastel.kit.edu/uncertainty/}} to conveniently calculate biometric uncertainty beyond the heuristic cases, based on BioQuake. With these tools, we aim at promoting widespread use of uncertainty reporting and thus, comparable research.
    
    \item We conduct a \textbf{comprehensive analysis of the uncertainty in biometric research.} (\cref{sec:ExperimentalSetup_lr}, \cref{sec:resultsUncertainty}.) We analyzed the uncertainty in the reported performance of state-of-the-art biometric solutions across 62 biometric datasets spanning 8 modalities (face, fingerprint, gait, iris, keystroke, eye movement, EEG, ECG). Our findings show that the performance of \textcolor{black}{many systems} might significantly deviate from actual results, highlighting the urgency of reporting error uncertainties in biometric research.
\end{enumerate}

\section{Related Work}
\label{sec:RelatedWork}
In biometric recognition, research results are commonly reported either solely by error rates \cite{sluganovic2018analysis,fallahi2023brainnet,melzi2023ecg,bidgoly2022towards} or, occasionally, including the standard error to indicate variations \cite{arias2023performance,maiorana2021learning}. However, confidence intervals ensure a more reliable quantification of performance. For error rates, a confidence interval provides an error range that likely contains the true error with a specific probability.
Therefore, the size of the confidence range reflects the uncertainty given a predefined confidence level. Despite the importance of uncertainty estimation, the process of calculating confidence intervals can be challenging \cite{jolliffe2007uncertainty}, and reporting uncertainty is rarely considered a mandatory step for publishing research in biometrics.

\subsection{Confidence Estimation For Biometric Recognition}
Approximating confidence intervals is typically achieved through two main approaches: Distribution-based methods~\cite{agresti2012categorical,garthwaite2002statistical} and the Bootstrap method~\cite{diciccio1996bootstrap,carpenter2000bootstrap}. Distribution-based methods require assumptions about the data, such as sample size, population variance (known or unknown), or the underlying distribution. In contrast, the Bootstrap method is a computational technique that operates without these assumptions. It estimates confidence intervals by repeatedly sampling from the data with replacement, facilitating the direct calculation of intervals from the sample itself. Although not as robust as distribution-based methods, the Bootstrap method is often advantageous when the data distribution is unknown and computational overhead is negligible.

For biometric verification tasks, several works have explored quantifying uncertainty by adapting confidence intervals. Predominantly, these works employ a distribution-based approach due to their reliability and lower computational demands. Initial research by Shen et al. \cite{shen1997evaluation} focused on parameter estimation from repeated Bernoulli experiments to establish confidence intervals, simplifying the approximation of the binomial distribution to a normal distribution. Following this methodology, Veres et al. \cite{veres2006enough} utilized the Chernoff bound to approximate the binomial distribution with a normal distribution. Moreover, Li et al. \cite{li2012test} introduced confidence elasticity as a metric to evaluate the trade-off between sample size and the precision of evaluation metrics again under the assumption of normal distribution. On the other hand, employing the bootstrap approach, Bolle et al. \cite{bolle2004error} developed the subsets bootstrap method to calculate the confidence intervals for the Receiver Operating Characteristic (ROC) curves in biometric recognition systems. Additionally, Dass et al. (2006) \cite{dass2006validating} merged both methodologies, utilizing a multivariate joint distribution through a parametric family of Gaussian copulas and the bootstrap method to estimate confidence intervals for ROC curves.

\subsection{Limitations of Previous Works}
Studies on estimating confidence intervals for biometric verification results have primarily two limitations. First, the proposed methods are complicated to apply without specialized statistical knowledge, creating a burden to use these methods in practice. Secondly, many confidence estimation methods make strong assumptions regarding their error distribution, such as assuming a normal distribution of the samples and uniform sample sizes across subjects, which are not given in real systems and thus, are mostly not valid. 
To address these issues, we introduce an uncertainty estimation framework that builds on more realistic assumptions and is easy to apply to arbitrary biometric verification tasks.
This aims to encourage the use of uncertainties in biometrics and thus, to promote the development of more reliable biometric solutions and datasets.

\section{Methodology}
\label{sec:Methodology}
Uncertainty estimation is important to quantify the reliability of a biometric verification system. To quantify the uncertainty, we initially formulate a confidence interval in the context of biometric verification, named BioQuake, and establish straightforward rules that researchers can easily apply to calculate the required number of comparisons to reach a desired uncertainty or to estimate the uncertainty based on the number of comparison scores. We then define uncertainty classes to facilitate easy comparison of the uncertainty between biometric solutions and finally, we discuss the upper and lower bounds of BioQuake.

\subsection{Introducing BioQuake $(\delta)$}
\label{sec:MethodologyRU}
 Evaluating biometric verification systems requires the computation of comparison scores of genuine and imposter pairs to calculate FMR and FNMR. To formulate the FMR and FNMR based on the underlying distributions of comparison scores, the distributions of genuine and imposter comparison scores are denoted as $P_{G}$ and $P_{I}$, respectively.
Assuming a decision threshold of $t$, the FMR and FNMR are defined as
\begin{align}
    FMR &= \int_{t}^{\infty} P_{I}(s)\,ds =: p_{FM} \\
    FNMR &= \int_{-\infty}^{t} P_{G}(s)\, ds =: p_{FNM}.
\end{align}
The FMR can be described as the probability $p_{FM}$ that an imposter comparison is falsely matched as a genuine comparison.
Similarly, the FNMR can be defined as the probability $p_{FNM}$ that a genuine comparison is falsely considered as a non-match (imposter).

In the following, we assume drawing $N$ samples independently from a score distribution $P_O$, where $O\in{I,G}$ is either an imposter ($I$) or genuine ($G$) score distribution. 
The probability of a faulty decision based on a score $s$ drawn from this distribution is denoted as $p$.
For $O=I$ imposter, this happens for $s\geq t$ (a false match) and for $O=G$ genuine, this happens for $s<t$ (a false non-match).
Consequently, the probability of observing $n$ wrong matching decisions from the $N$ i.i.d. samples drawn from $P_O$ can be modeled by a binomial distribution:
\begin{align}
    p(X=n)= {N \choose n} \, p^{n}\, (1-p)^{N-n}
\end{align}
To calculate the acceptance region that the observed error rate $p$ is reasonable with confidence of $1-\alpha$, we can compute the lower and higher boundary $[n_L, n_H]$ by finding the highest $n_L$ and the lowest $n_H$ that fullfills
\begin{align}
    P(X<n_L)&=\sum_{n=0}^{n_L} p(X=n) \leq \dfrac{\alpha}{2} \\
    P(X>n_H)&= 1 - \sum_{n=0}^{n_H} p(X=n) \leq \dfrac{\alpha}{2} .
\end{align}
By setting the acceptance region $[n_L, n_H]$ in relation to the number of observed samples $N$, we can derive an uncertainty range of the error rates 
\begin{align}
    \left[\frac{n_L}{N}, \frac{n_H}{N}\right] \hat{=} \left[\textit{FMR}_L, \textit{FMR}_H\right] \hat{=} \left[\textit{FNMR}_L, \textit{FNMR}_H \right],
\end{align}
i.e. FMR if $O=I$ or FNMR if $O=G$.
Assuming that the observed error rate $\frac{n}{N}$ is centered in this region, we can compute the uncertainty $\Delta$ of the error rate as
\begin{align}
    \Delta  = \dfrac{n_H-n_L}{2N}.
    \label{eq:AR}
\end{align}
This means if we observe, an FNMR $=2\%$ and a $\Delta=1\%$, the true FNMR value might lay within the region of $2\% \pm 1\%$ with a probability of $1-\alpha$.

To make the comparison of the uncertainties comparable for different error rates, we further introduce the term \textit{BioQuake} $(\delta)$ which is defined as 
\begin{align}
    \delta = \dfrac{\Delta}{p}.
    \label{eq:RU}
\end{align}
BioQuake is defined as the ratio of the uncertainty $\Delta$ and the observed error rate $p$,  and it describes the uncertainty with respect to the verification error.
For instance, $\delta=0.1$ means that the uncertainty is 10\% of the observed error rate, indicating a more precise estimate. In contrast, $\delta=1$  refers to an uncertainty as large as the error rate itself, suggesting a much less reliable estimate. This will allow us later to comprehensively compare the uncertainties across different biometric algorithms and modalities.

\subsection{Determining BioQuake Visually}
\label{sec:MethodologyParameterAnalysis}

\begin{figure*}
    \centering
    \subfloat[90\% Confidence \label{fig1}]{%
       \includegraphics[width=0.33\textwidth]{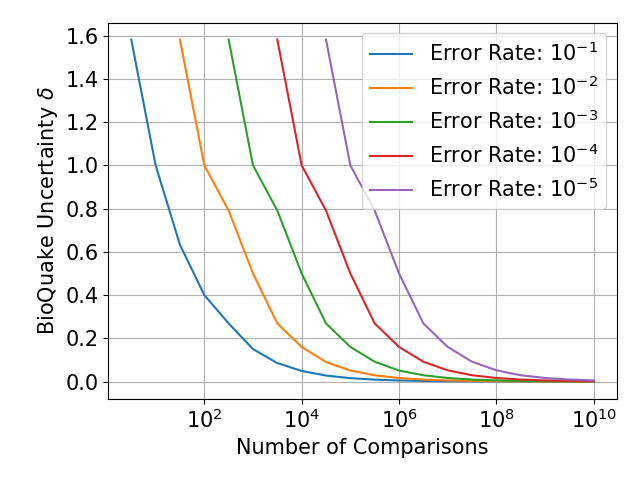}}
    \subfloat[95\% Confidence \label{fig1}]{%
       \includegraphics[width=0.33\textwidth]{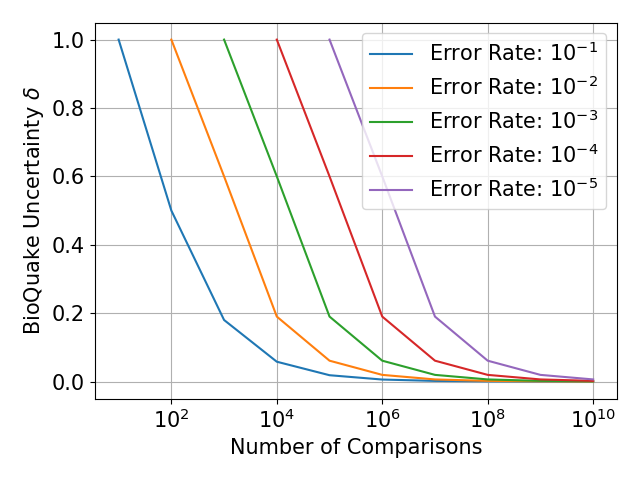}}
    \subfloat[99\% Confidence \label{fig1}]{%
       \includegraphics[width=0.33\textwidth]{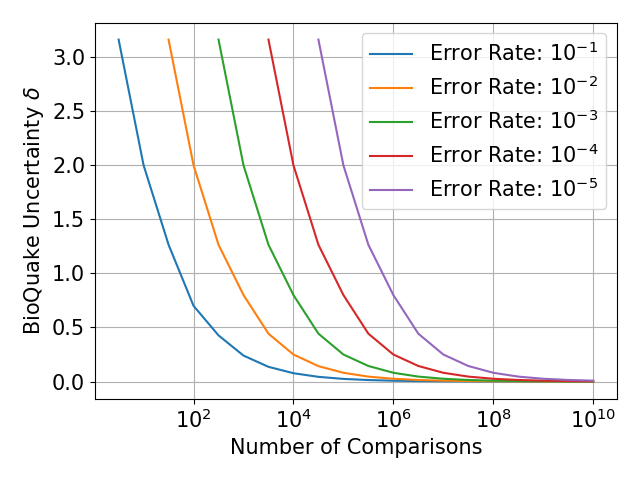}}
    \caption{\textbf{Visualization of Bioquake Uncertainty at Different Confidence Levels -} The relationship between the observed error rate (FMR/FNMR) and the required number of comparisons to achieve a specific BioQuake uncertainty is displayed for the confidence levels of 90\% ($\alpha=10\%$), 95\% ($\alpha=5\%$), and 99\% ($\alpha=1\%$).}
    \label{fig:3}
\end{figure*}

To provide better intuition, we visualize the relationship between error rate, and the minimum number of comparisons required to achieve a specific BioQuake uncertainty for three different confidence levels of 90\% ($\alpha=10\%$), 95\% ($\alpha=5\%$), and 99\% ($\alpha=1\%$).
These figures demonstrate that a lower error rate requires a larger number of samples to maintain a specific BioQuake rate, and achieving a higher confidence level requires a larger sample size (Figure \ref{fig:3}). 
Given the number of comparisons and an observed error rate, these figures allow for determining the uncertainty of the reported error rate observation.

\begin{figure}
    \centering
    \includegraphics[width=0.49\textwidth]{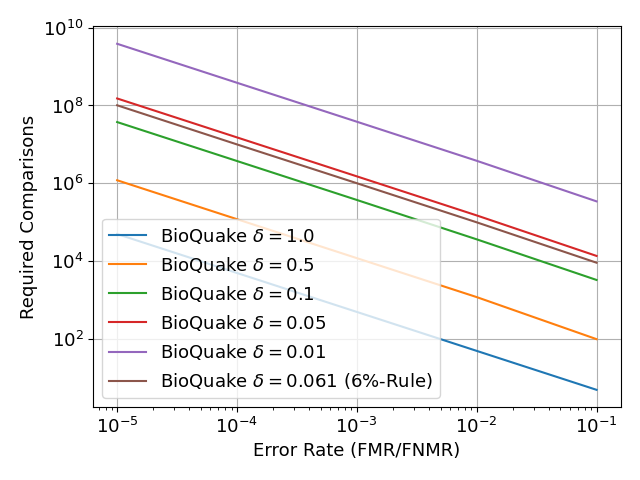}
    \caption{
    \textbf{Visualization of BioQuake Principles -} The relationship between the observed error rate (FMR/FNMR) and the required number of comparisons to achieve a specific BioQuake uncertainty $\delta$ is displayed (for $\alpha=0.05$). Based on this, the required number of comparisons for a specific uncertainty, as well as the significance of measurement, could be determined.}
    \label{fig:2}
\end{figure}

For a confidence level of 95\% ($\alpha=5\%$), Figure \ref{fig:2} illustrates the relationship between the observed error rate and the required number of comparisons to achieve a specific BioQuake uncertainty.
This figure enables estimation of the required number of comparisons to reach a desired uncertainty level and indicates the error rate range within which reporting remains reliable.

\subsection{BioQuake Rules}
\label{sec:MethodologyRules}
Next, we want to define some easily applicable rules for estimating the uncertainty of biometric verification tasks and the number of comparisons needed for significant performance reporting.
These rules rely on the linear relationship between the error rate $p$ and the number of comparisons $N$ for a fixed BioQuake uncertainty ($\delta$) level seen in Figure \ref{fig:2}. 
Based on this, we define some straightforward rules of thumb to can be easily applied.
The 1\% and 10\% rules ensure that the measured error only differs by 1\% and 10\% from the true error.
We included a 6\% rule since it leads to an easy-to-remember rule of thumb.

\begin{tcolorbox}[breakable, boxrule=0pt] \textbf{The 1\% Rule ($\delta=0.01$)} – To measure an Error Rate (ER), such as the False Match Rate (FMR) or False Non-Match Rate (FNMR), and ensure with 95\% confidence that the true error differs by no more than 1\% from the measured error, a minimum of $\textit{RC} = \frac{3.83\times 10^4}{\textit{ER}}$ comparisons is required. \end{tcolorbox}

\begin{tcolorbox}[breakable, boxrule=0pt] \textbf{The 6\% Rule ($\delta=0.061$)} – To measure an Error Rate (ER), such as the False Match Rate (FMR) or False Non-Match Rate (FNMR), and ensure with 95\% confidence that the true error differs by no more than 6.1\% from the measured error, a minimum of $\textit{RC} = \frac{10^3}{\textit{ER}}$ comparisons is required. \end{tcolorbox}

\begin{tcolorbox}[breakable, boxrule=0pt] \textbf{The 10\% Rule ($\delta=0.1$)} – To measure an Error Rate (ER), such as the False Match Rate (FMR) or False Non-Match Rate (FNMR), and ensure with 95\% confidence that the true error differs by no more than 10\% from the measured error, a minimum of $\textit{RC} = \frac{3.7\times 10^2}{\textit{ER}}$ comparisons is required. \end{tcolorbox}

For dealing with FMR, the Required Comparisons (RC) reflect the number of imposter comparisons needed.
For FNMR, the RC indicates the number of genuine comparisons.

For example, if an error for a FMR of $10^{-3}$ should be measured, according to the 6\% - Rule, you need at least $10^6$ imposter comparisons to ensure that the measured error only differs by around 6\% of the true one.
Similarly, the 10\% rule states that to measure an error for an FNMR of $10^{-3}$, at least $3.7 \times 10^5$ genuine samples are needed to ensure that the measured error deviates by no more than 10\% from the true error. Additionally, to achieve an error measurement with a 1\% uncertainty for the same FNMR, the number of genuine comparisons must increase by a factor of 100, requiring at least $3.83 \times 10^7$ comparisons.

\subsection{Certainty Classes}
\label{sec:Methodology_CertaintyClasses}
To facilitate understanding and comparison of reported uncertainty across different biometric solutions and modalities, we introduce an uncertainty classification as follows:

\begin{itemize}
    \item \begin{listbox}[colback=highlightgreen!50] \textbf{Class A+ (Optimal):} $\delta < 0.01$ \end{listbox}
    \item \begin{listbox}[colback=highlightblue!50] \textbf{Class A (Excellent):} $0.01 \leq \delta < 0.05$ \end{listbox}
    \item \begin{listbox}[colback=highlightper!50] \textbf{Class B (Very Good):} $0.05 \leq \delta < 0.10$ \end{listbox}    
    \item \begin{listbox}[colback=highlightyellow!50] \textbf{Class C (Good):} $0.10 \leq \delta < 0.30$ \end{listbox}
    \item \begin{listbox}[colback=highlightorange!50] \textbf{Class D (Fair):} $0.30 \leq \delta < 0.50$ \end{listbox}
    \item \begin{listbox}[colback=highlightbrown!50] \textbf{Class E (Poor):} $0.50 \leq \delta < 1.00$ \end{listbox}
    \item \begin{listbox}[colback=highlightred!50] \textbf{Class F (Unacceptable):} $\delta > 1$ \end{listbox}
\end{itemize}

Classes A+ represent cases with minimal uncertainty, indicating high reliability in the reported results, with a drift potential of less than 1\% for true error. Classes A and B represent low uncertainty, showing only minor variability that does not substantially impact the trustworthiness of the results. Classes C and D denote moderate uncertainty, where the results have reduced reliability but remain within acceptable limits for most applications. Class E signifies significant risk, with uncertainty approaching or equaling the magnitude of the reported result, necessitating caution in data interpretation. Class F is designated for extreme cases where the uncertainty exceeds the reported result, indicating highly unreliable data that could mislead decision-making processes.

\label{sec:MethodologyClasses}

\subsection{Upper and Lower Bounds of Uncertainty}
Since it is important to understand the range of possible BioQuake values when evaluating biometric solutions, we define the upper and lower bounds of uncertainty $(\Delta)$ based on Equation \ref{eq:AR}. Given $n_{L/H} \in \{0,1,\dots,N\}$, the bounds for the error rate uncertainty are:
\begin{align}
    \Delta_{\max} = \dfrac{1}{2}, \qquad
    \Delta_{\min} = \dfrac{1}{2N}
\end{align}
The upper limit \(\Delta_{\max} = \frac{1}{2}\) is logical, as it accounts for the full range of potential error rates in both directions. The lower limit \(\Delta_{\min} = \frac{1}{2N}\) represents the smallest uncertainty measurable with this approach. Given these bounds for \(\Delta\), we can determine the bounds for BioQuake \((\delta)\)
\begin{align}
    \delta_{\max} = \infty, \qquad
    \delta_{\min} = \dfrac{1}{2N}.
\end{align}
The upper limit \((\delta_{\max})\) occurs under conditions of minimal error \(p\) combined with \(\Delta_{\max}\), and the lower limit \((\delta_{\min})\) corresponds to \(\Delta_{\min}\) with an error equal to 100\%. This means that, with a very low error rate based on a small number of observations, the relative uncertainty can become extremely large, potentially approaching infinity. In contrast, if an error rate of 1 is observed, the uncertainty decreases proportionally with the number of evaluations. In practice, most situations lie between these extremes, indicating that increasing the number of observations reduces uncertainty. However, reporting a very low error rate with reasonable certainty requires a much larger sample size. This is because accurately estimating a small error rate demands enough observations for the error to occur sufficiently for a reliable estimate. This issue becomes even more critical in the context of the FMR, where minimizing the probability of attacker access necessitates reporting a very low error rate to ensure system robustness.

\section{Experimental setup} 
\label{sec:ExperimentalSetup}
To validate the effectiveness and relevance of the \textit{BioQuake} metric, we conducted two experiments. First, we demonstrate the correctness of our approach empirically. Second, we applied our metric to assess the uncertainty of errors in state-of-the-art biometric solutions on various benchmarks and datasets to investigate the significance of the solutions tested on this data.

\subsection{Setup A: Empirical Correctness Analysis of BioQuake} 
\label{sec:ExperimentalSetup_ee}
To empirically evaluate the theoretical uncertainty calculated by the proposed \textit{BioQuake} approach, we design an experiment including three large facial datasets and four well-known face recognition models.
These datasets and models were selected due to their extensive development, rigorous testing, and significant representation in the literature. Below, we provide a detailed description of the experimental setup.

\textbf{Datasets}: We utilized three facial recognition datasets to assess the uncertainty of FMR and FNMR empirically: the ColorFERET dataset~\cite{phillips1998feret,phillips2000feret}, the Labeled Faces in the Wild (LFW) dataset~\cite{huang2008labeled}, and the Adience dataset~\cite{eidinger2014age}. The ColorFERET dataset consists of over 11,000 images captured under controlled conditions, featuring systematic variations in pose, expression, and lighting to maintain consistent testing environments. In contrast, the LFW dataset comprises over 13,000 faces sourced from the web under less controlled but high-quality images.
While LFW often refers to a benchmark based on these images with only 6k predefined comparisons, we utilize LFW (full), referring to the usage of all images against each other. The Adience dataset contains over 26,000 everyday images and focuses on unconstrained images with often low image quality. Collectively, these datasets facilitate comprehensive testing of our uncertainty assessment algorithm across various facial recognition scenarios.

\textbf{Metrics}: Following the ISO standard \cite{ISO19795-1:2021} for measuring biometric system performance, the primary metrics discussed are the False Non-Match Rate (FNMR) and the False Match Rate (FMR). 
The FNMR quantifies the likelihood that the biometric system will incorrectly reject an attempt by an authorized user to gain access. Conversely, the FMR measures the probability that the system will incorrectly accept an unauthorized user.

\textbf{Utilized Models}: To extract face templates and thus, calculate comparison scores and FMR/FNMR, we employed four well-known pre-trained face recognition models. Each model offers unique features and methodologies for vector representation essential for our analysis. FaceNet~\cite{schroff2015facenet}, pioneered by Google, employs a triplet loss function, which trains the network to map facial images into Euclidean space where the distances equate to facial similarity. This technique enhances face recognition by reducing distances between similar faces and increasing those between dissimilar ones, thus improving accuracy in identity verification. ArcFace~\cite{deng2019arcface} builds upon FaceNet's methodology by incorporating an additive angular margin loss, which increases the geometric margin, improving the model's discriminative capacity. This adjustment aids in achieving tighter intra-class compactness and enhanced inter-class separability.
MagFace~\cite{meng2021magface} introduces a magnitude-aware loss function that varies the loss contribution according to the embedding's magnitude, which correlates with the facial image's quality; higher-quality images result in larger magnitudes. This strategy encourages the model to prioritize higher-quality images while continuing to learn from those of lower quality. 
Finally, QMagFace \cite{terhorst2023qmagface} integrates the quality information of the face samples in the comparison process to enhance its recognition performance under unconstrained circumstances.
For the experiments, we used the (ResNet-100/iResNet-101) models from the official repositories of the authors to extract the face templates. Comparing the two templates was done based on the standard cosine similarity.

\textbf{Method}: To investigate empirical uncertainty, extracted face templates from each face image and computed the cosine similarity between every distinct pair of images within each dataset, generating genuine scores for images from the same subject and impostor scores for pairs from different subjects. Then, for each dataset and model, the comparisons are randomly sampled with sizes of 0.1, 0.01, and 0.001 of the overall sample size (denoted as \textit{frac}), ten times, and then computed the FMR and FNMR.  
The standard deviation of these sample means was calculated to estimate the Standard Error of the Mean (SEM). This estimated SEM was then multiplied by 1.96 to determine the margin of error at a 95\% confidence level ~\cite{lee2015standard,altman2005standard}, representing the empirical uncertainty.
Finally, BioQuake, with the same 95\% confidence level, was calculated as the theoretical uncertainty predicted by our method to assess the similarity between empirical and theoretical uncertainty.

\subsection{Setup B: Uncertainty Analysis of Existing Datasets}
\label{sec:ExperimentalSetup_lr}
To explore uncertainty issues in current biometric verification research, we investigated 62 biometric datasets and benchmarks by analyzing state-of-the-art performances reported from works in leading conferences and journals in the field. Our review covered a range of eight biometric modalities, including facial recognition and fingerprints as widely used physical biometrics, as well as gait, iris, Electrocardiogram (ECG), Electroencephalogram (EEG), eye-tracking analysis, and keystroke dynamics as emerging behavioral biometrics. We selected papers only from Q1 and Q2 journals and A+ and A conferences. Although, IJCB is not ranked in the Scientific Journal Rankings~\footnote{\url{https://www.scimagojr.com/journalrank.php}} and Core ranking~\footnote{\url{https://portal.core.edu.au/conf-ranks/}}, in the community it is considered as one of the three top biometric conferences along IAPR ICB and IEEE BTAS.

For each paper, we extracted the number of impostor and genuine comparisons used to evaluate error rates. Then, we calculated the \textit{BioQuake} for each error type with a 95\% confidence interval. Additionally, we estimated the minimum reportable error rate, setting the delta value at 6.1\% (calculated using $ER = \frac{10^3}{NC}$, where NC is the number of available comparisons), to elucidate the limitations of current error reporting methodologies and motivate future studies to improve error reporting via \textit{BioQuake}.

\begin{table*}[h!]
\footnotesize
\renewcommand{\arraystretch}{1.0}
\setlength{\tabcolsep}{2.0pt}
    \centering
    \footnotesize
    \caption{ \textbf{Uncertainty Analysis on Biometric Recognition Datasets} - \textcolor{black}{ Performance of state-of-the-art biometric systems from top-tier venues using 62 datasets across 8 modalities. For each dataset, we report the number of recorded data collection sessions (\#S), identities (IDs), imposter (Imp) and genuine (Gen) comparisons, and performance in terms of FNMR and FMR. Datasets marked with * where used in different configurations, described in the referenced publications. For each performance metric, we computed the BioQuake uncertainty ($\delta$) and the number of minimal errors measured for the 6\% rule. Legend: Uncertainty is color-coded as \colorbox{highlightgreen!50}{Optimal}, \colorbox{highlightblue!50}{Excellent}, \colorbox{highlightper!50}{Very Good}, \colorbox{highlightyellow!50}{Good}, \colorbox{highlightorange!50}{Fair}, \colorbox{highlightbrown!50}{Poor}, and \colorbox{highlightred!50}{Unnacceptable}; NA = Not Available, ; Private = Dataset not Publicly Available (Though information about identity comparisons is contained in the publication)} 
}

    \label{tab:RelatedWork}
\begin{tabular}{clcccccclccccccc}
\Xhline{2\arrayrulewidth}
 & \multicolumn{5}{c}{Dataset} & \multicolumn{4}{c}{Publication} & \multicolumn{2}{c}{FNMR} & \multicolumn{2}{c}{FMR} & \multicolumn{2}{c}{Min. Error for $\delta=6.1\%$}\\
 
\cmidrule(rl){2-6} \cmidrule(rl){7-10} \cmidrule(rl){11-12} \cmidrule(rl){13-14} \cmidrule(rl){15-16}
   & Dataset&\#S& IDs & Imp & Gen & Publ.& Year & Venue & Rank&   FNMR [\%]& $\delta_{FNMR}$ & FMR [\%]& $\delta_{FMR}$ &  FNMR  &  FMR\\
   \hline

\parbox[t]{2mm}{\multirow{13}{*}{\rotatebox[origin=c]{90}{ECG}}}

& Private~\cite{kang2016ecg} & 1 & 28 & 3.9K & 140 & \cite{kang2016ecg} & 2016 & IEEE SPL & Q1 & 1.90 & \cellcolor{highlightbrown!50}{0.93984} & 5.20 & \cellcolor{highlightyellow!50}{0.13067} & $\geq1$ & $3 * 10^{-1}$ \\

& ECG-ID*~\cite{lugovaya2005biometric} & $>$1 & 90 & 4M & 45K & \cite{chu2019ecg} & 2019 & IEEE Access & Q1 & 2.00 & \cellcolor{highlightper!50}{0.06444} & 2.00 & \cellcolor{highlightgreen!50}{0.006849} & $2 * 10^{-2}$ & $2 * 10^{-4}$ \\
& E-HOL~\cite{moody2001impact} & 1 & 202 & 4.2M & 45.8K & \cite{labati2019deep} & 2019 & PRL & Q1 & 2.15 & \cellcolor{highlightper!50}{0.06144} & 2.15 & \cellcolor{highlightgreen!50}{0.00644} & $2 * 10^{-2}$ & $2 * 10^{-4}$ \\
& MIT-BIH~\cite{moody2001impact} & $>$1 & 47 & 1M & 23.5K & \cite{chu2019ecg} & 2019 & IEEE Access & Q1 & 4.74 & \cellcolor{highlightper!50}{0.05688} & 4.74 & \cellcolor{highlightgreen!50}{0.00875} & $4 * 10^{-2}$ & $1 * 10^{-3}$ \\
& PTB*~\cite{bousseljot1995nutzung} & 5 & 52 & 1.3M & 26K & \cite{chu2019ecg} & 2019 & IEEE Access & Q1 & 0.59 & \cellcolor{highlightyellow!50}{0.15319} & 0.59 & \cellcolor{highlightblue!50}{0.02229} & $4 * 10^{-2}$ & $8 * 10^{-4}$ \\
& CYBHi*~\cite{da2014check} & 2 & 63 & 315 & 63 & \cite{melzi2023ecg} & 2023 & IEEE Access & Q1 & 6.98 & \cellcolor{highlightbrown!50}{0.79592} & 6.98 & \cellcolor{highlightorange!50}{0.36385} & $\geq1$ & $\geq1$ \\
& CYBHi*~\cite{da2014check} & 1 & 63 & 945 & 189 & \cite{melzi2023ecg} & 2023 & IEEE Access & Q1 & 5.44 & \cellcolor{highlightbrown!50}{0.53493} & 5.44 & \cellcolor{highlightyellow!50}{0.2626} & $\geq1$ & $\geq1$ \\
& ECG-ID*~\cite{lugovaya2005biometric} & 1 & 89 & 1.3K & 267 & \cite{melzi2023ecg} & 2023 & IEEE Access & Q1 & 1.52 & \cellcolor{highlightbrown!50}{0.7392} & 1.52 & \cellcolor{highlightorange!50}{0.40485} & $\geq1$ & $8 * 10^{-1}$ \\
& ECG-ID*~\cite{lugovaya2005biometric} & $>$1 & 89 & 445 & 89 & \cite{melzi2023ecg} & 2023 & IEEE Access & Q1 & 0.26 & \cellcolor{highlightred!50}{$>1$} & 0.26 & \cellcolor{highlightred!50}{$>1$} & $\geq1$ & $\geq1$ \\
& In-house*~\cite{melzi2023ecg} & 1 & 55.9K & 77.6K & 15.5K & \cite{melzi2023ecg} & 2023 & IEEE Access & Q1 & 1.28 & \cellcolor{highlightyellow!50}{0.13608} & 1.28 & \cellcolor{highlightper!50}{0.06141} & $6 * 10^{-2}$ & $1 * 10^{-2}$ \\
& In-house*~\cite{melzi2023ecg} & 2 & 26K & 25.8K & 5.1K & \cite{melzi2023ecg} & 2023 & IEEE Access & Q1 & 1.97 & \cellcolor{highlightyellow!50}{0.18413} & 1.97 & \cellcolor{highlightper!50}{0.08558} & $2 * 10^{-1}$ & $4 * 10^{-2}$ \\
& PTB*~\cite{bousseljot1995nutzung} & 1 & 113 & 1.7K & 339 & \cite{melzi2023ecg} & 2023 & IEEE Access & Q1 & 0.14 & \cellcolor{highlightred!50}{$>1$} & 0.14 & \cellcolor{highlightred!50}{$>1$} & $\geq1$ & $6 * 10^{-1}$ \\
& PTB*~\cite{bousseljot1995nutzung} & 5 & 113 & 565 & 113 & \cite{melzi2023ecg} & 2023 & IEEE Access & Q1 & 2.06 & \cellcolor{highlightred!50}{$>1$} & 2.06 & \cellcolor{highlightbrown!50}{0.5155} & $\geq1$ & $\geq1$ \\
\hline
\parbox[t]{2mm}{\multirow{5}{*}{\rotatebox[origin=c]{90}{EEG}}}& Private~\cite{maiorana2021learning} & 5 & 45 & 239.5M & 17.1M & \cite{maiorana2021learning} & 2021 & PRL & Q1 & 4.80 & \cellcolor{highlightgreen!50}{0.00211} & 4.80 & \cellcolor{highlightgreen!50}{0.00056} & $6 * 10^{-5}$ & $4 * 10^{-6}$ \\
& EEG Motor~\cite{goldberger2000physiobank} & 1 & 109 & 468.9M & 13M & \cite{bidgoly2022towards} & 2022 & PRL & Q1 & 1.96 & \cellcolor{highlightgreen!50}{0.0039} & 1.96 & \cellcolor{highlightgreen!50}{0.00065} & $8 * 10^{-5}$ & $2 * 10^{-6}$ \\
& bi2015a~\cite{korczowski2019brain} & 1 & 40 & 42.4K & 10.6K & \cite{fallahi2023brainnet} & 2023 & PerCom & A+ & 0.21 & \cellcolor{highlightorange!50}{0.38134} & 1.00 & \cellcolor{highlightper!50}{0.09303} & $9 * 10^{-2}$ & $2 * 10^{-2}$ \\
& Private~\cite{arias2023performance} & 1 & 49 & 36.5K & 0.8K & \cite{arias2023performance} & 2023 & ACM TOPS & Q1 & 8.50 & \cellcolor{highlightyellow!50}{0.22058} & 8.50 & \cellcolor{highlightblue!50}{0.03336} & $\geq1$ & $3 * 10^{-2}$ \\
& Private~\cite{tajdini2023brainwave} & 1 & 50 & 122.5K & 2.5K & \cite{tajdini2023brainwave} & 2023 & C\&S & Q1 & 0.52 & \cellcolor{highlightorange!50}{0.48148} & 0.52 & \cellcolor{highlightper!50}{0.07482} & $4 * 10^{-1}$ & $8 * 10^{-3}$ \\
\hline
\parbox[t]{2mm}{\multirow{7}{*}{\rotatebox[origin=c]{90}{Eye-tracking}}}& Private~\cite{sluganovic2018analysis} & 4 & 30 & 0.5K & 1K & \cite{sluganovic2018analysis} & 2018 & ACM TOPS & Q1 & 6.30 & \cellcolor{highlightyellow!50}{0.23809} & 6.30 & \cellcolor{highlightorange!50}{0.30103} & $\geq1$ & $\geq1$ \\
& Private~\cite{sluganovic2018analysis} & 4 & 30 & 1M & 1K & \cite{sluganovic2018analysis} & 2018 & ACM TOPS & Q1 & 1.37 & \cellcolor{highlightbrown!50}{0.50043} & 0.06 & \cellcolor{highlightper!50}{0.07833} & $\geq1$ & $1 * 10^{-3}$ \\
& Private~\cite{zhang2018continuous} & 2 & 30 & 1K & 90 & \cite{zhang2018continuous} & 2018 & ACM IMWUT & Q2 & 6.90 & \cellcolor{highlightbrown!50}{0.64412} & 6.90 & \cellcolor{highlightyellow!50}{0.21739} & $\geq1$ & $\geq1$ \\
& Private~\cite{eberz201928} & 1 & 22 & 6.9K & 0.3K & \cite{eberz201928} & 2019 & CCS & A+ & 1.88 & \cellcolor{highlightbrown!50}{0.70921} & 1.88 & \cellcolor{highlightyellow!50}{0.16574} & $\geq1$ & $1 * 10^{-1}$ \\
& GazeBase*~\cite{griffith2021gazebase} & 9 & 322 & 1.49M & 98K & \cite{yin2022effective} & 2022 & Sensors & Q1 & 10.62 & \cellcolor{highlightblue!50}{0.01811} & 10.62 & \cellcolor{highlightgreen!50}{0.00465} & $1 * 10^{-2}$ & $7 * 10^{-4}$ \\
& GazeBase*~\cite{griffith2021gazebase} & 9 & 322 & 96k & 5.9K & \cite{yin2022effective} & 2022 & Sensors & Q1 & 5.25 & \cellcolor{highlightyellow!50}{0.10653} & 5.25 & \cellcolor{highlightblue!50}{0.02678} & $2 * 10^{-1}$ & $1 * 10^{-2}$ \\
& GazeBase*~\cite{griffith2021gazebase} & 9 & 322 & 20K & 20K & \cite{lohr2022eye} & 2022 & IEEE TIFS & Q1 & 5.09 & \cellcolor{highlightper!50}{0.05893} & 0.01 & \cellcolor{highlightred!50}{$>1$} & $5 * 10^{-2}$ & $5 * 10^{-2}$ \\
\hline
\parbox[t]{2mm}{\multirow{15}{*}{\rotatebox[origin=c]{90}{Face}}}& LFW ~\cite{huang2008labeled} & NA & 1.6K & 3k & 3k & \cite{schroff2015facenet} & 2015 & CVPR & A+ & 0.37 & \cellcolor{highlightbrown!50}{0.54054} & 0.37 & \cellcolor{highlightbrown!50}{0.54054} & $3 * 10^{-1}$ & $3 * 10^{-1}$ \\
& LEO\_LS~\cite{best2017longitudinal} & NA & 5K & 5.5M & 26K & \cite{best2017longitudinal} & 2018 & IEEE TPAMI & Q1 & 0.34 & \cellcolor{highlightyellow!50}{0.19796} & 0.10 & \cellcolor{highlightblue!50}{0.02636} & $4 * 10^{-2}$ & $2 * 10^{-4}$ \\
& PCSO\_LS~\cite{best2017longitudinal} & NA & 18K & 11.1B & 129K & \cite{best2017longitudinal} & 2018 & IEEE TPAMI & Q1 & 2.17 & \cellcolor{highlightblue!50}{0.03654} & 0.10 & \cellcolor{highlightgreen!50}{0.00587} & $8 * 10^{-3}$ & $9 * 10^{-8}$ \\
& IJB-B~\cite{whitelam2017iarpa} & NA & 1.8K & 8M & 10K & \cite{deng2019arcface} & 2019 & CVPR & A+ & 5.80 & \cellcolor{highlightper!50}{0.07758} & 0.01 & \cellcolor{highlightper!50}{0.06874} & $1 * 10^{-1}$ & $1 * 10^{-4}$ \\
& IJB-C~\cite{maze2018iarpa} & NA & 3.5K & 15.6M & 19.5K & \cite{deng2019arcface} & 2019 & CVPR & A+ & 4.40 & \cellcolor{highlightper!50}{0.06526} & 0.01 & \cellcolor{highlightblue!50}{0.04935} & $5 * 10^{-2}$ & $6 * 10^{-5}$ \\
& Trillion P.~\cite{deng2019arcface} & NA & 5.7K & 330.1B & 11M & \cite{deng2019arcface} & 2019 & CVPR & A+ & 18.44 & \cellcolor{highlightgreen!50}{0.00124} & $10^{-7}$ & \cellcolor{highlightyellow!50}{0.10601} & $9 * 10^{-5}$ & $3 * 10^{-9}$ \\
& RFW~\cite{wang2019racial} & NA & 11.4K & 50M & 14K & \cite{wang2020mitigating} & 2020 & CVPR & A+ & 4.81 & \cellcolor{highlightper!50}{0.0735} & 4.81 & \cellcolor{highlightgreen!50}{0.00123} & $7 * 10^{-2}$ & $2 * 10^{-5}$ \\
& CALFW~\cite{zheng2017cross} & NA & 1.6K & 3k & 3k & \cite{meng2021magface} & 2021 & CVPR & A+ & 7.23 & \cellcolor{highlightyellow!50}{0.12448} & 7.23 & \cellcolor{highlightyellow!50}{0.12448} & $3 * 10^{-1}$ & $3 * 10^{-1}$ \\
& IJB-B~\cite{whitelam2017iarpa} & NA & 1.8K & 8M & 10K & \cite{kim2022adaface} & 2022 & CVPR & A+ & 3.97 & \cellcolor{highlightper!50}{0.09571} & 3.97 & \cellcolor{highlightgreen!50}{0.0034} & $1 * 10^{-1}$ & $1 * 10^{-4}$ \\
& IJB-C~\cite{maze2018iarpa} & NA & 3.5K & 15.6M & 19.5K & \cite{kim2022adaface} & 2022 & CVPR & A+ & 2.41 & \cellcolor{highlightper!50}{0.0883} & 2.41 & \cellcolor{highlightgreen!50}{0.00314} & $5 * 10^{-2}$ & $6 * 10^{-5}$ \\
& LFW ~\cite{huang2008labeled} & NA & 1.6K & 3k & 3k & \cite{kim2022adaface} & 2022 & CVPR & A+ & 0.20 & \cellcolor{highlightbrown!50}{0.66666} & 0.20 & \cellcolor{highlightbrown!50}{0.66666} & $3 * 10^{-1}$ & $3 * 10^{-1}$ \\
& AgeDB~\cite{moschoglou2017agedb} & NA & 0.5K & 3k & 3k & \cite{terhorst2023qmagface} & 2023 & CVPR & A+ & 1.50 & \cellcolor{highlightyellow!50}{0.28888} & 1.50 & \cellcolor{highlightyellow!50}{0.28888} & $3 * 10^{-1}$ & $3 * 10^{-1}$ \\
& LFW (full) ~\cite{huang2008labeled} & NA & 3k & 87M & 240K & Our & 2024 & TIFS & Q1 & 0.04 & \cellcolor{highlightyellow!50}{0.19791} & 0.04 & \cellcolor{highlightblue!50}{0.01048} & $4 * 10^{-3}$ & $1 * 10^{-5}$ \\
& ColorFERET (full) \cite{phillips2000feret} & NA & 1k & 63M & 95K & Our & 2024 & TIFS & Q1 & 1.83 & \cellcolor{highlightblue!50}{0.0463} & 1.83 & \cellcolor{highlightgreen!50}{0.0018} & $1 * 10^{-2}$ & $2 * 10^{-5}$ \\
& Adience (full) ~\cite{eidinger2014age} & NA & 3k & 186M & 682K & Our & 2024 & TIFS & Q1 & 2.21 & \cellcolor{highlightblue!50}{0.01575} & 2.21 & \cellcolor{highlightgreen!50}{0.00019} & $1 * 10^{-3}$ & $5 * 10^{-6}$ \\
\hline
\parbox[t]{2mm}{\multirow{6}{*}{\rotatebox[origin=c]{90}{Fingerprint}}}& FVC2002~\cite{maio2002fvc2002} & NA & 110 & 4.9K & 2.8K & \cite{jea2005minutia} & 2005 & PR & Q1 & 1.57 & \cellcolor{highlightyellow!50}{0.27297} & 1.57 & \cellcolor{highlightyellow!50}{0.21231} & $4 * 10^{-1}$ & $2 * 10^{-1}$ \\
& FVC2002~\cite{maio2002fvc2002} & NA & 110 & 4.9K & 2.8K & \cite{jiang2005orientation} & 2005 & IEEE TSP & Q1 & 1.93 & \cellcolor{highlightyellow!50}{0.25906} & 1.93 & \cellcolor{highlightyellow!50}{0.19562} & $4 * 10^{-1}$ & $2 * 10^{-1}$ \\
& FVC2000~\cite{maio2002fvc2000} & NA & 100 & 9.9K & 0.6K & \cite{uz2009minutiae} & 2009 & CVIU & Q1 & 4.68 & \cellcolor{highlightorange!50}{0.35612} & 0.10 & \cellcolor{highlightbrown!50}{0.60606} & $\geq1$ & $1 * 10^{-1}$ \\
& FVC2006~\cite{cappelli2007fingerprint} & NA & 140 & 9.7k & 9.2k & \cite{cappelli2010minutia} & 2010 & IEEE TPAMI & Q1 & 0.15 & \cellcolor{highlightbrown!50}{0.50505} & 0.15 & \cellcolor{highlightorange!50}{0.47961} & $1 * 10^{-1}$ & $1 * 10^{-1}$ \\
& FVC2004~\cite{maio2004fvc2004} & NA & 100 & 4.9K & 2.8K & \cite{cao2011fingerprint} & 2011 & IEEE IJCB & NA & 2.18 & \cellcolor{highlightyellow!50}{0.24574} & 0.10 & \cellcolor{highlightbrown!50}{0.80808} & $4 * 10^{-1}$ & $2 * 10^{-1}$ \\
& LFIW~\cite{liu2024latent} & NA & 132 & 173.6M & 118.1K & \cite{liu2024latent} & 2024 & IEEE TIFS & Q1 & 22.82 & \cellcolor{highlightblue!50}{0.01048} & 22.82 & \cellcolor{highlightgreen!50}{0.00027} & $8 * 10^{-3}$ & $6 * 10^{-6}$ \\
\hline
\parbox[t]{2mm}{\multirow{4}{*}{\rotatebox[origin=c]{90}{Gait}}}& Private~\cite{xu2018keh} & 20 & 2 & 1K & 1K & \cite{xu2018keh} & 2019 & IEEE TMC & Q1 & 12.10 & \cellcolor{highlightyellow!50}{0.16528} & 12.10 & \cellcolor{highlightyellow!50}{0.16528} & $\geq1$ & $\geq1$ \\
& Private~\cite{xu2018keh} & 20 & 2 & 1K & 2K & \cite{xu2018keh} & 2019 & IEEE TMC & Q1 & 6.00 & \cellcolor{highlightyellow!50}{0.23333} & 6.00 & \cellcolor{highlightyellow!50}{0.16666} & $5 * 10^{-1}$ & $\geq1$ \\
& Private~\cite{zhang2020deepkey} & 7 & 3 & 1K & 1K & \cite{zhang2020deepkey} & 2020 & TIST & Q1 & 0.39 & \cellcolor{highlightbrown!50}{0.76923} & 0.39 & \cellcolor{highlightbrown!50}{0.76923} & $\geq1$ & $\geq1$ \\
& Private~\cite{ji2021one} & 9 & 1 & 14.4K & 144K & \cite{ji2021one} & 2021 & IEEE SPL & Q1 & 22.20 & \cellcolor{highlightblue!50}{0.03034} & 22.2 & \cellcolor{highlightgreen!50}{0.00966} & $7 * 10^{-3}$ & $7 * 10^{-2}$ \\
\hline
\parbox[t]{2mm}{\multirow{4}{*}{\rotatebox[origin=c]{90}{Iris}}}& CASIA-IrisV1~\cite{casiairisv1} & NA & 108 & 2.3K & 40.4K & \cite{chin2006high} & 2006 & CVIU & Q1 & 0.25 & \cellcolor{highlightbrown!50}{0.69565} & 0.25 & \cellcolor{highlightyellow!50}{0.18811} & $2 * 10^{-2}$ & $4 * 10^{-1}$ \\
& CASIA-IrisV3*~\cite{casiairisv3} & NA & 396 & 2.6K & 373.7K & \cite{lai2017cancellable} & 2017 & PR & Q1 & 0.54 & \cellcolor{highlightorange!50}{0.49857} & 0.54 & \cellcolor{highlightblue!50}{0.04311} & $3 * 10^{-3}$ & $4 * 10^{-1}$ \\
& CASIA-IrisV3*~\cite{casiairisv3} & NA & 700 & 0.7K & 0.7K & \cite{asaker2021novel} & 2021 & MTA & Q1 & 1.93 & \cellcolor{highlightorange!50}{0.48112} & 1.93 & \cellcolor{highlightorange!50}{0.48112} & $\geq1$ & $\geq1$ \\
& CASIA-IrisV3*~\cite{casiairisv3} & NA & 411 & 0.4K & 505K & \cite{kausar2021iris} & 2021 & EIJ & Q1 & 2.00 & \cellcolor{highlightbrown!50}{0.625} & 0.80 & \cellcolor{highlightblue!50}{0.03069} & $2 * 10^{-3}$ & $\geq1$ \\
\hline
\parbox[t]{2mm}{\multirow{8}{*}{\rotatebox[origin=c]{90}{Keystroke}}}& Private~\cite{hwang2009keystroke} & 1 & 25 & 1.3K & 0.8K & \cite{hwang2009keystroke} & 2009 & C\&S & Q1 & 4.00 & \cellcolor{highlightorange!50}{0.31249} & 4.00 & \cellcolor{highlightyellow!50}{0.25} & $\geq1$ & $8 * 10^{-1}$ \\
& Giot~\cite{giot2009greyc} & 5 & 100 & 33K & 5.5K & \cite{giot2011unconstrained} & 2011 & C\&S & Q1 & 6.69 & \cellcolor{highlightper!50}{0.09783} & 6.69 & \cellcolor{highlightblue!50}{0.04008} & $2 * 10^{-1}$ & $3 * 10^{-2}$ \\
& Sheng~\cite{sheng2005parallel} & NA & 33 & 27.9K & 0.9K & \cite{balagani2011discriminability} & 2011 & PRL & Q1 & 4.13 & \cellcolor{highlightyellow!50}{0.29593} & 4.13 & \cellcolor{highlightper!50}{0.05597} & $\geq1$ & $4 * 10^{-2}$ \\
& Private~\cite{locklear2014continuous} & 2 & 486 & 70.7M & 145.8K & \cite{locklear2014continuous} & 2014 & IEEE IJCB & NA & 4.55 & \cellcolor{highlightblue!50}{0.02344} & 4.55 & \cellcolor{highlightgreen!50}{0.00106} & $7 * 10^{-3}$ & $1 * 10^{-5}$ \\
& Buffalo~\cite{sun2016shared} & 3 & 157 & 39.5M & 260.1K & \cite{lu2020continuous} & 2020 & C\&S & Q1 & 2.36 & \cellcolor{highlightblue!50}{0.02468} & 2.36 & \cellcolor{highlightgreen!50}{0.002} & $4 * 10^{-3}$ & $3 * 10^{-5}$ \\
& Clarkson~\cite{murphy2017shared} & NA & 103 & 131.3M & 1.3M & \cite{lu2020continuous} & 2020 & C\&S & Q1 & 5.97 & \cellcolor{highlightgreen!50}{0.00067} & 5.97 & \cellcolor{highlightgreen!50}{0.00681} & $8 * 10^{-4}$ & $8 * 10^{-6}$ \\
& Kim~\cite{kim2020freely} & 1 & 50 & 24.5K & 0.5K & \cite{kim2020freely} & 2020 & PR & Q1 & 0.02 & \cellcolor{highlightred!50}{$>1$} & 0.02 & \cellcolor{highlightbrown!50}{0.81632} & $\geq1$ & $4 * 10^{-2}$ \\
& Private~\cite{stylios2021bioprivacy} & 16 & 39 & 1.5K & 38 & \cite{stylios2021bioprivacy} & 2021 & ESORICS & A & 0.02 & \cellcolor{highlightred!50}{$>1$} & 0.02 & \cellcolor{highlightred!50}{$>1$} & $\geq1$ & $7 * 10^{-1}$ 

\\
\Xhline{2\arrayrulewidth}
\end{tabular}
\end{table*}


\section{Results}
\label{sec:Results}

\subsection{Empirical Correctness Analysis of BioQuake}
\label{sec:resultsCorrectness}
In this section, the correctness of the theoretical uncertainty measure BioQuake is analysed empircally on the three described large-scale face recognition datasets on four popular face recognition systems as described in Section \ref{sec:ExperimentalSetup_ee}.

\textbf{Correlation Analysis:} 
To investigate wether the proposed theoretical uncertainty measure BioQuake reflects the empirical uncertainty, the Pearson correlation between the theoretical and empirical uncertainty values are computed over different datasets and face recognition models.
In Table \ref{tab:CorrelationAnalysis}, these correlations are shown by computing the average correlation for each dataset-model combination.
The correlation values indicate a strong positive relationship between theoretical and empirical uncertainty across different models. Typically, a Pearson correlation greater than 0.7 indicates a strong positive linear relationship \cite{ratner2009correlation}. Thus, the high correlations observed confirms the effectiveness of our approach in estimating uncertainty across diverse datasets.

\begin{table}
    \centering
    \caption{\textbf{Correlation Analysis between the theoretical BioQuake and empirical Uncertainty} - Average Pearson correlation between theoretical (BioQuake) and empirical uncertainties across different models and datasets. The high correlation demonstrates that the proposed theoretical uncertainty highly reflects the empirical one.}
    \renewcommand{\arraystretch}{1.2}
    \begin{tabular}{llccc}
    \Xhline{2\arrayrulewidth}
        Model & Metric & ColorFERET & LFW & Adience \\
        \hline
        \multirow{2}{*}{QMagFace \cite{terhorst2023qmagface}} & FMR &0.9864 & 0.9893&0.9985 \\
                                  & FNMR &0.9878 & 0.9873&0.9776 \\
        \hline
        \multirow{2}{*}{MagFace \cite{meng2021magface}}  & FMR &0.9925 &0.9847 &0.9960 \\
                                  & FNMR &0.9889 &0.9905 &0.9715 \\
        \hline                          
        \multirow{2}{*}{ArcFace \cite{deng2019arcface}}  & FMR &0.9870 &0.9884 &0.9962 \\
                                  & FNMR &0.9771 &0.9808 &0.9856 \\
        \hline                          
        \multirow{2}{*}{FaceNet \cite{schroff2015facenet}}  & FMR &0.9945 &0.9828 &0.9974 \\
                                  & FNMR &0.9798 &0.9845 &0.9738 \\
       \Xhline{2\arrayrulewidth}
    \end{tabular}
    \label{tab:CorrelationAnalysis}
\end{table}

\begin{figure*}
    \centering
    \subfloat[QMagFace-ColorFERET]{\includegraphics[width=0.33\textwidth]{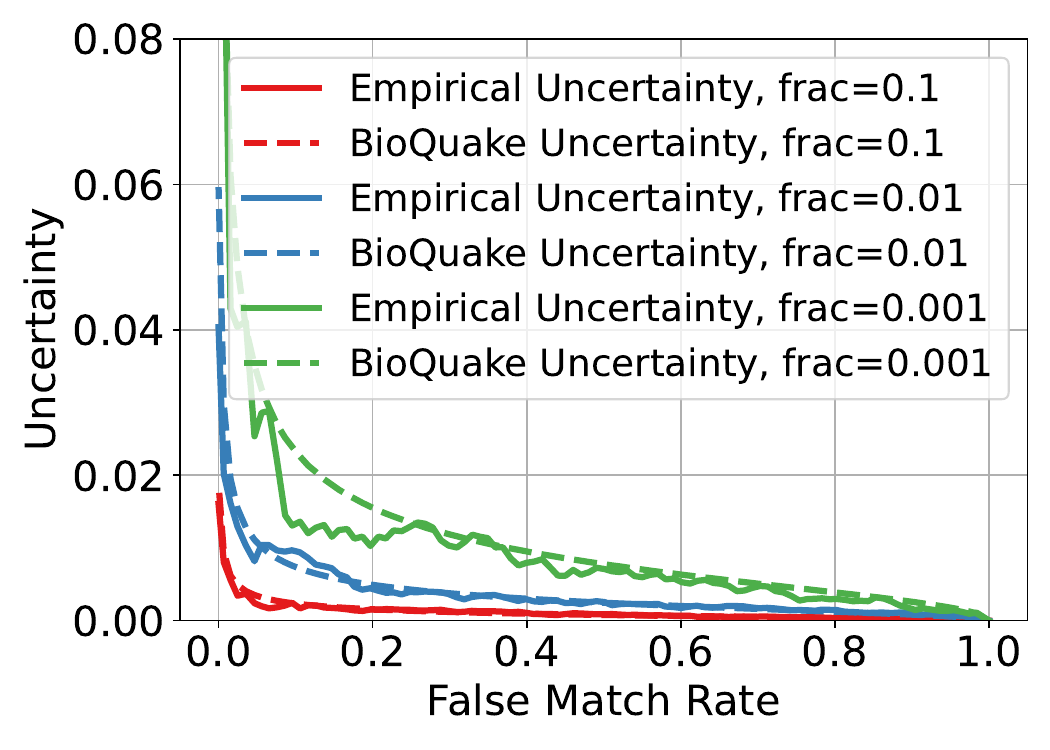}\label{fig:qmagface_colorferet}} \hfil
    \subfloat[QMagFace-LFW]{\includegraphics[width=0.33\textwidth]{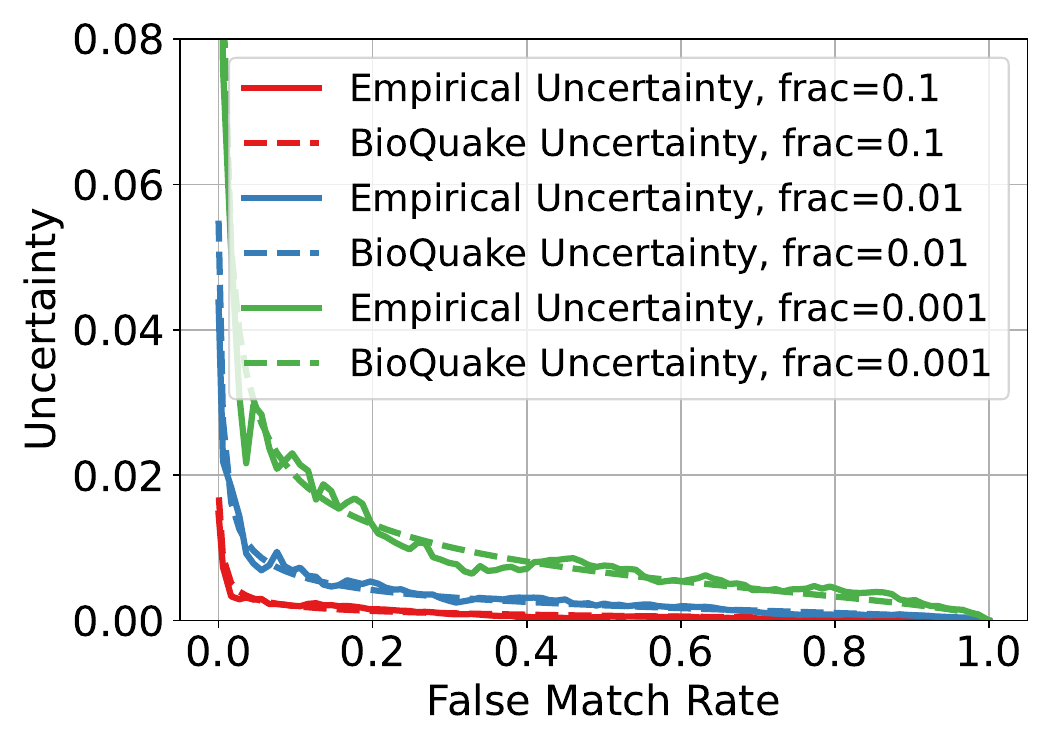}\label{fig:qmagface_lfw}} \hfil
    \subfloat[QMagFace-Adience]{\includegraphics[width=0.33\textwidth]{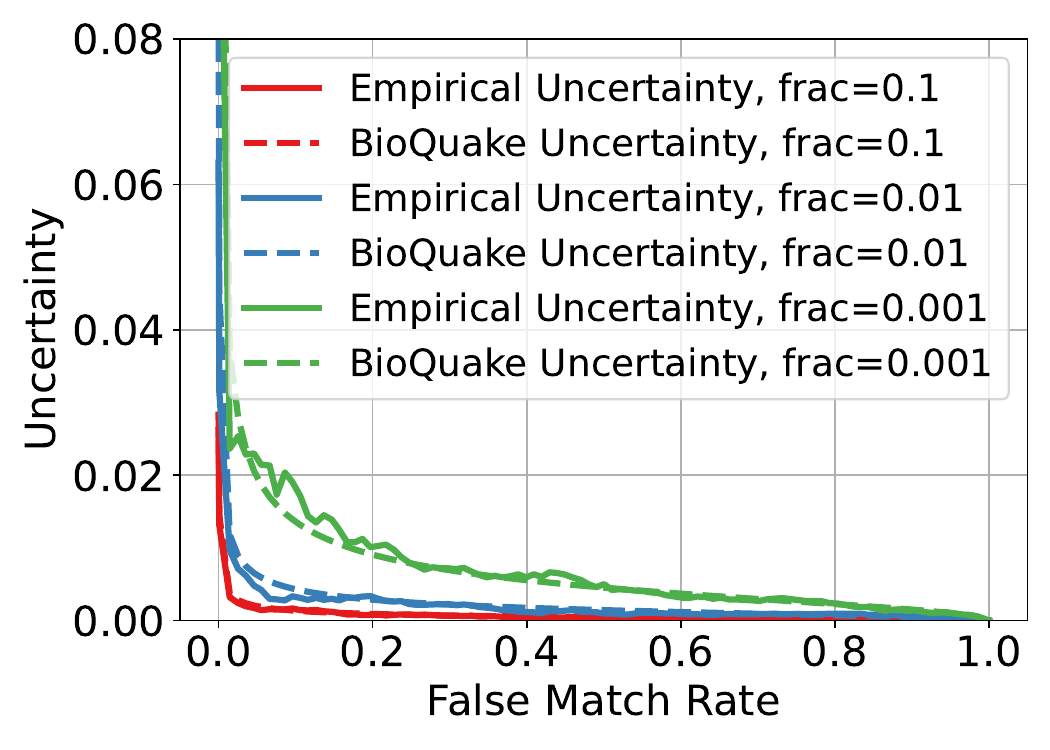}\label{fig:qmagface_adience}} \\
    
    \subfloat[MagFace-ColorFERET]{\includegraphics[width=0.33\textwidth]{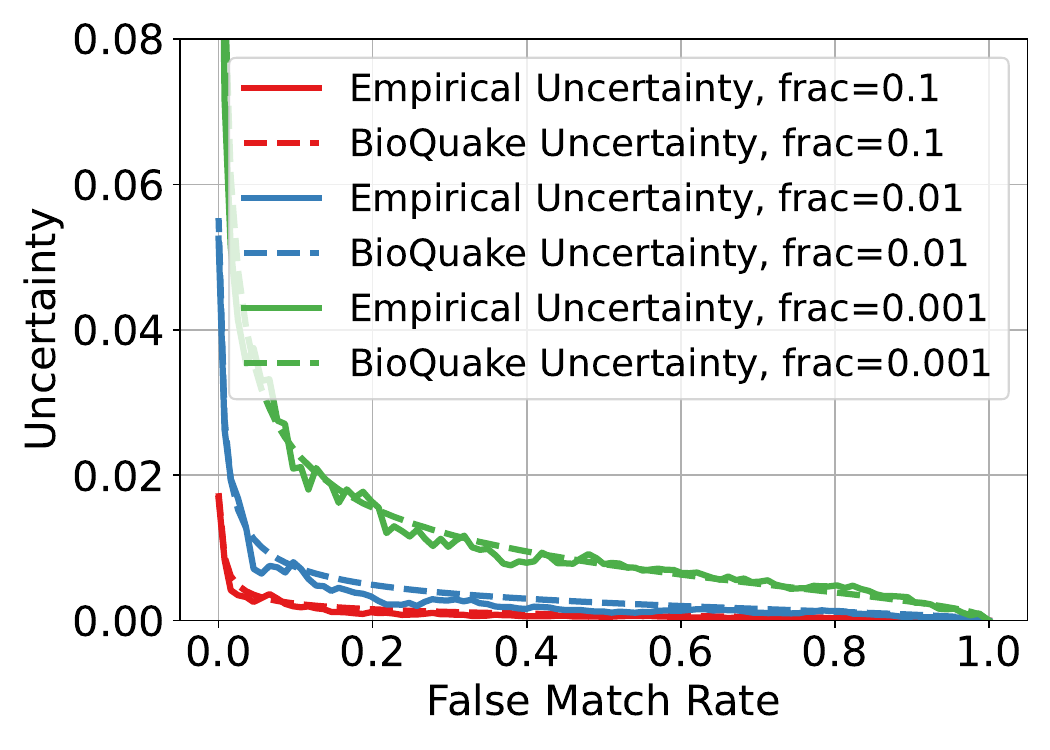}\label{fig:magface_colorferet}} \hfil
    \subfloat[MagFace-LFW]{\includegraphics[width=0.33\textwidth]{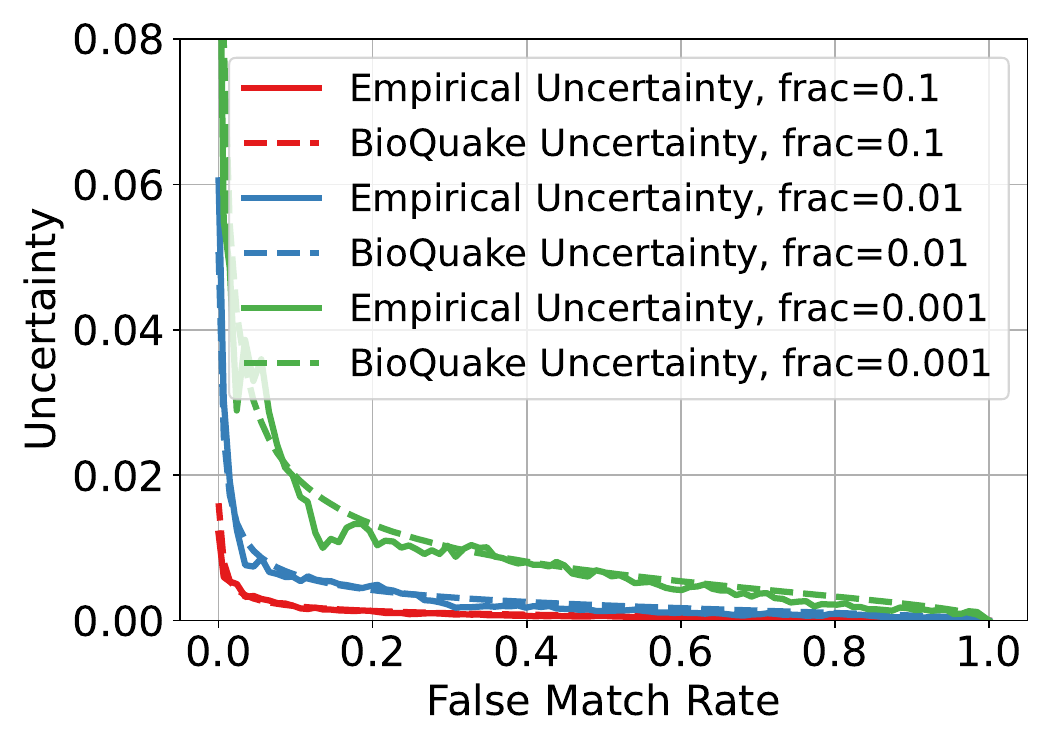}\label{fig:magface_lfw}} \hfil
    \subfloat[MagFace-Adience]{\includegraphics[width=0.33\textwidth]{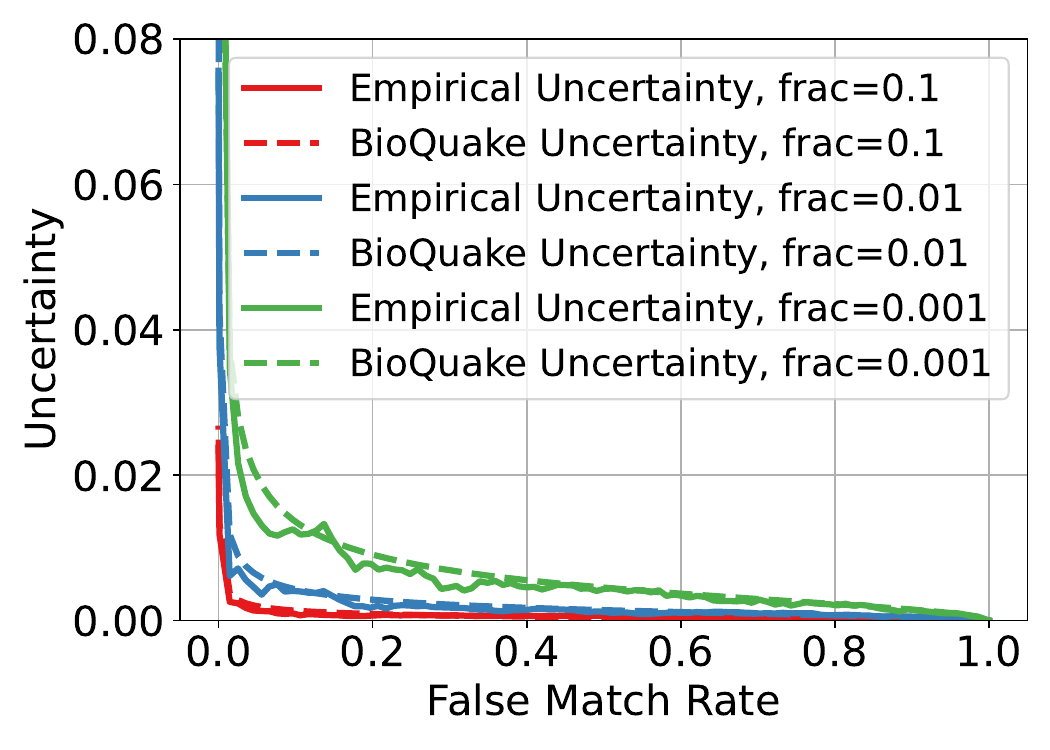}\label{fig:magface_adience}} \\

    \subfloat[ArcFace-ColorFERET]{\includegraphics[width=0.33\textwidth]{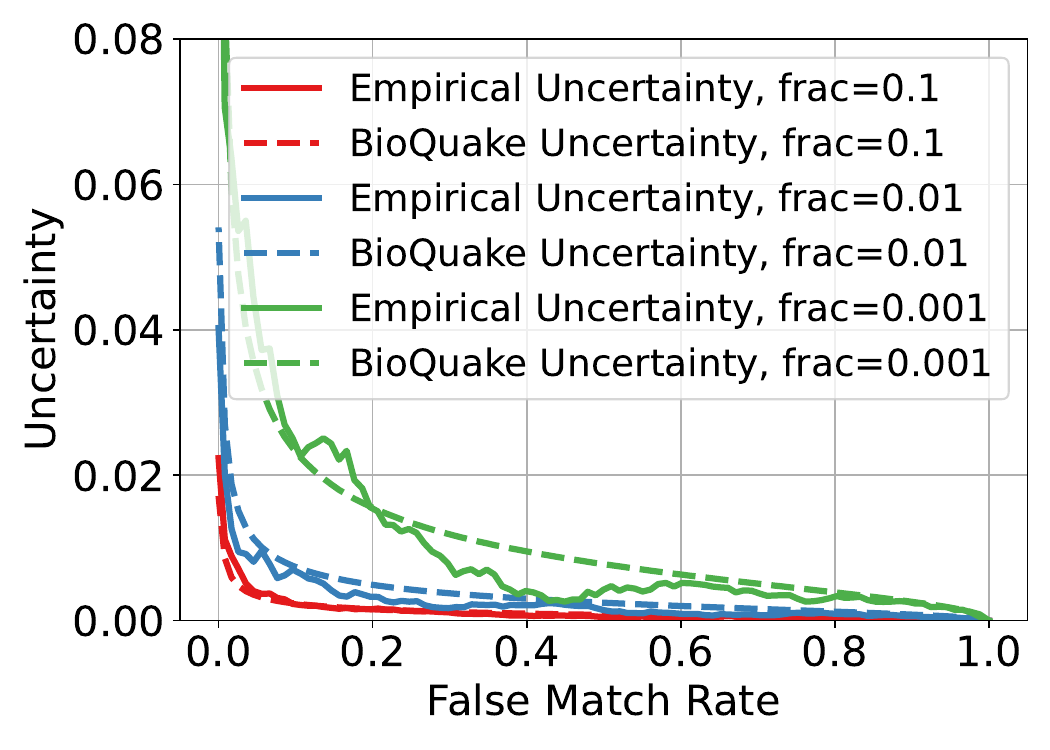}\label{fig:arcface_colorferet}} \hfil
    \subfloat[ArcFace-LFW]{\includegraphics[width=0.33\textwidth]{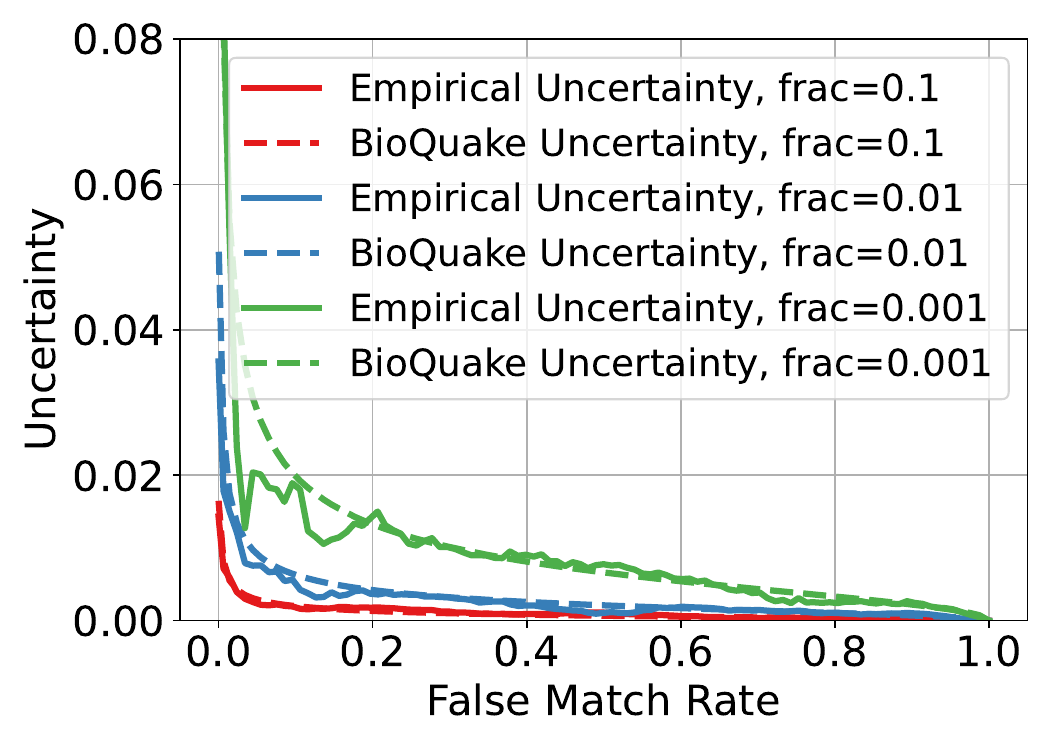}\label{fig:arcface_lfw}} \hfil
    \subfloat[ArcFace-Adience]{\includegraphics[width=0.33\textwidth]{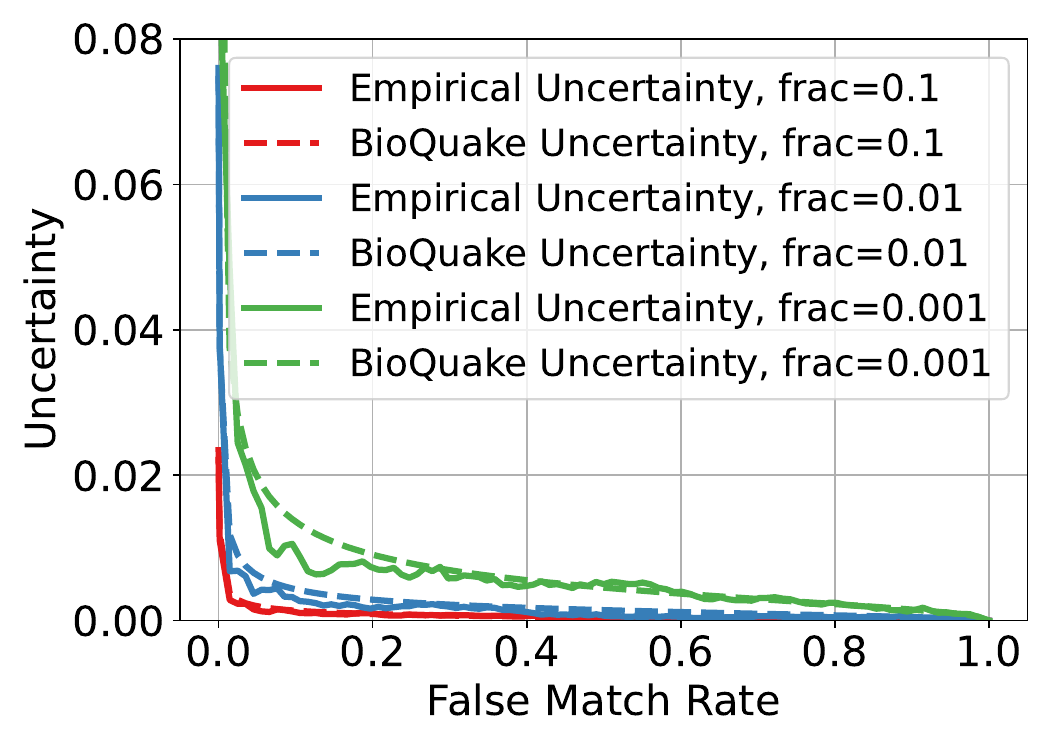}\label{fig:arcface_adience}} \\
    
    \subfloat[FaceNet-ColorFERET]{\includegraphics[width=0.33\textwidth]{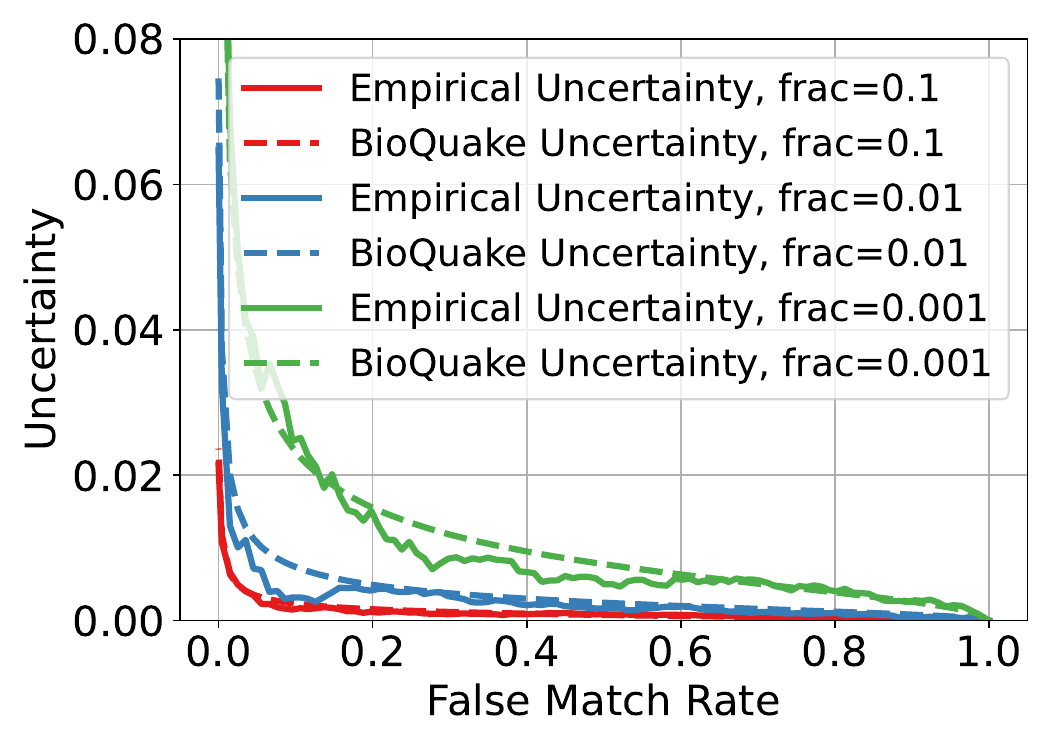}\label{fig:facenet_colorferet}} \hfil
    \subfloat[FaceNet-LFW]{\includegraphics[width=0.33\textwidth]{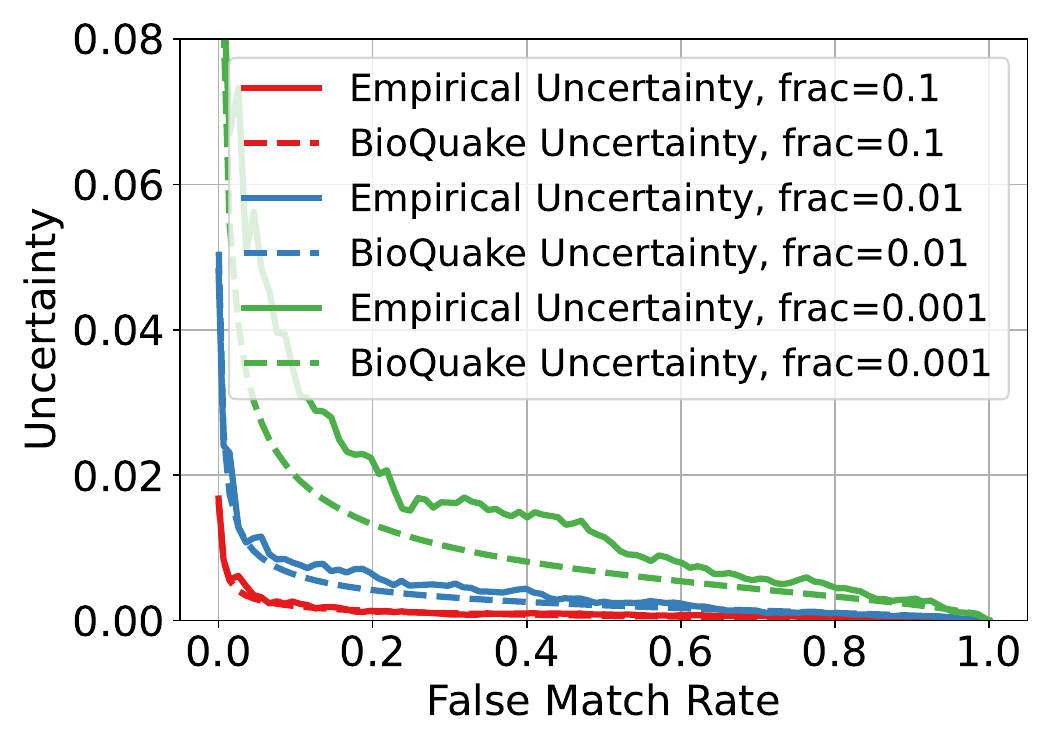}\label{fig:facenet_lfw}} \hfil
    \subfloat[FaceNet-Adience]{\includegraphics[width=0.33\textwidth]{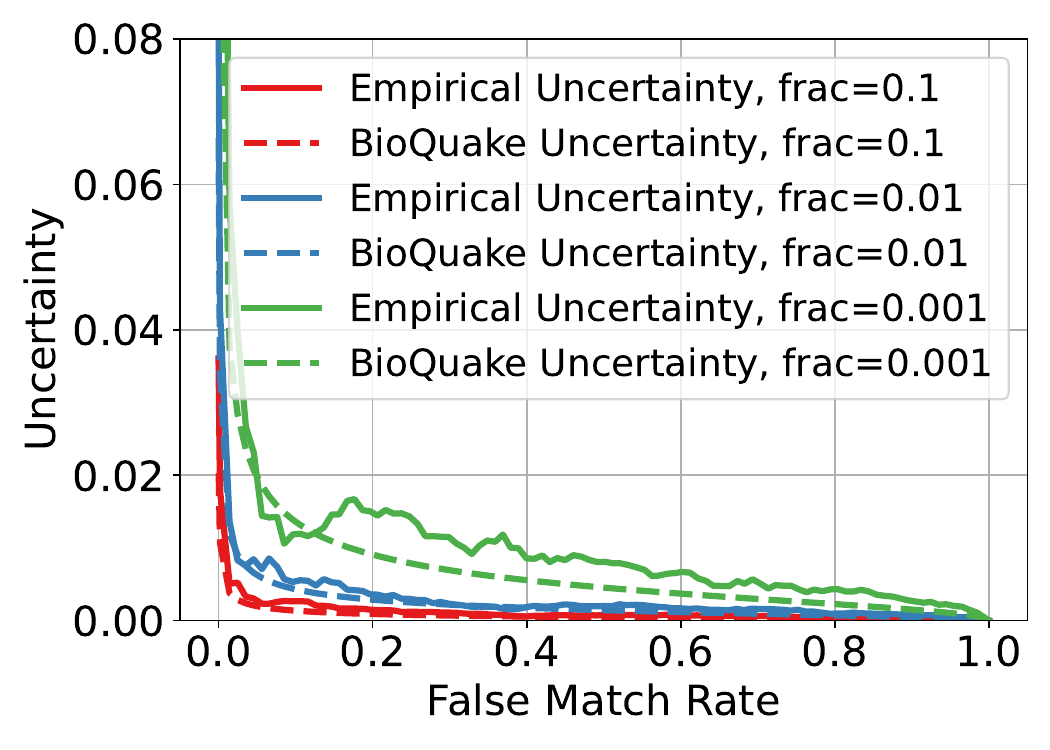}\label{fig:facenet_adience}} 

    \caption{\textbf{Comparative Analysis between the theoretical BioQuake and empircal Uncertainty for FMRs} - For different fractions (\textit{frac}) of the base datasets, the proposed theoretical BioQuake approach (dashed line) is compared against the empirical uncertainty (solid line) of FMRs on different model-dataset combinations. A high similarity between both approach is seen indicated the strong effectiveness of BioQuake in estimating the uncertainty.
    }
    \label{fig:fru1}
\end{figure*}

\begin{figure*}
    \centering

    \subfloat[QMagFace-ColorFERET]{\includegraphics[width=0.33\textwidth]{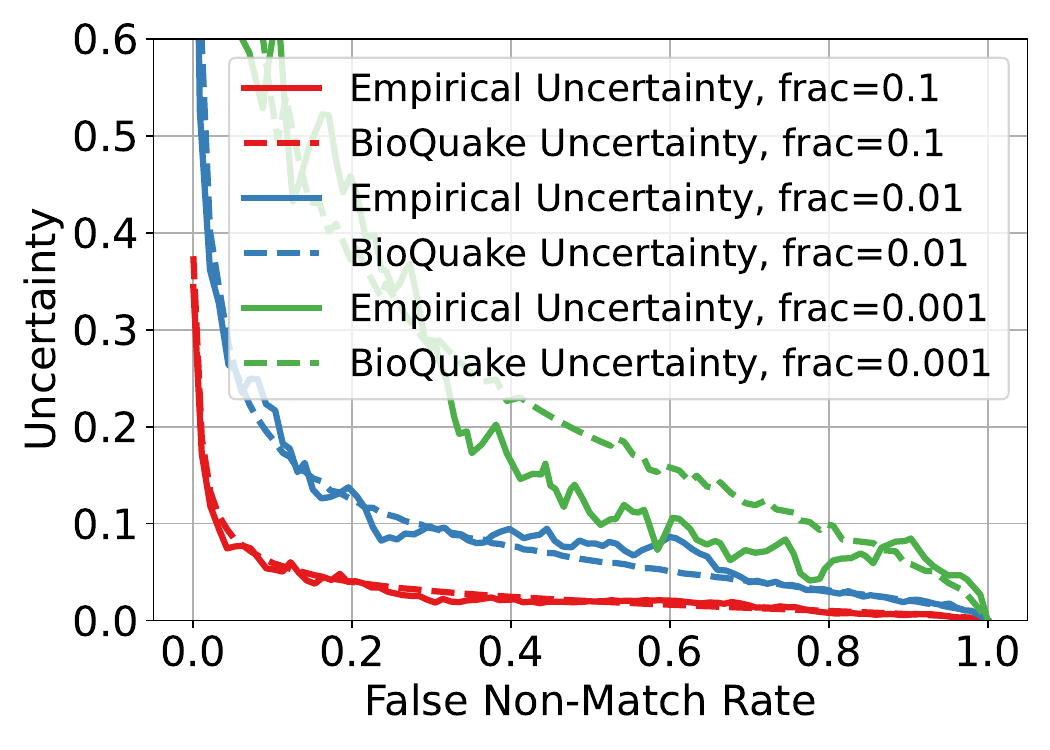}\label{fig:qmagface_colorferet}} \hfil
    \subfloat[QMagFace-LFW]{\includegraphics[width=0.33\textwidth]{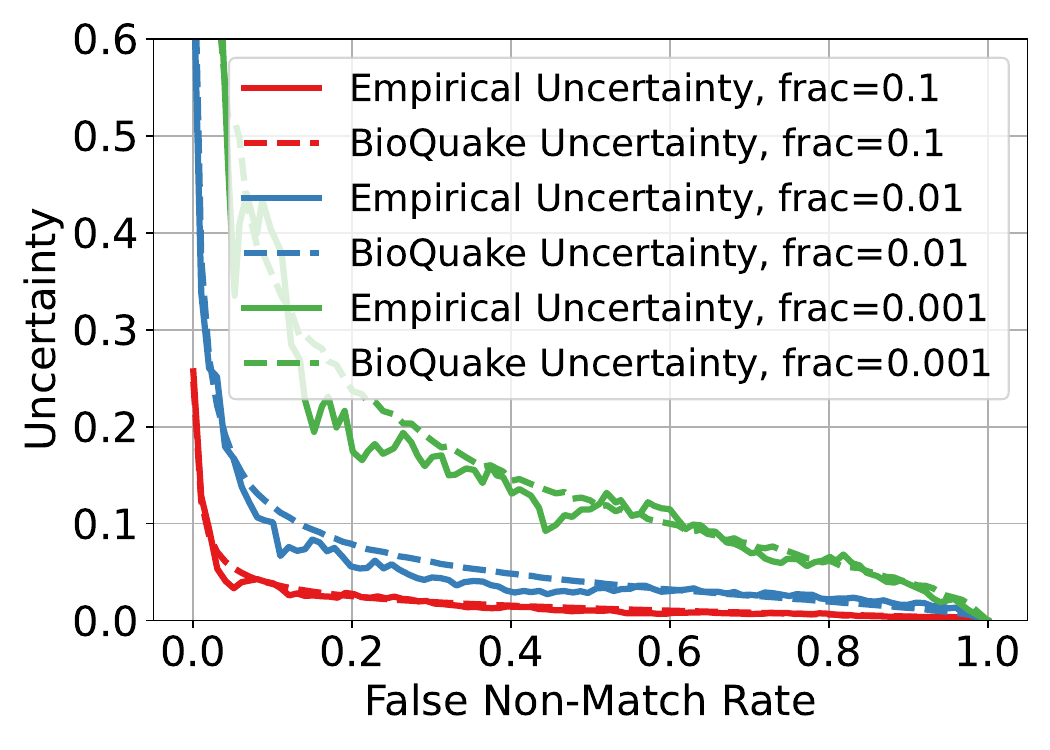}\label{fig:qmagface_lfw}} \hfil
    \subfloat[QMagFace-Adience]{\includegraphics[width=0.33\textwidth]{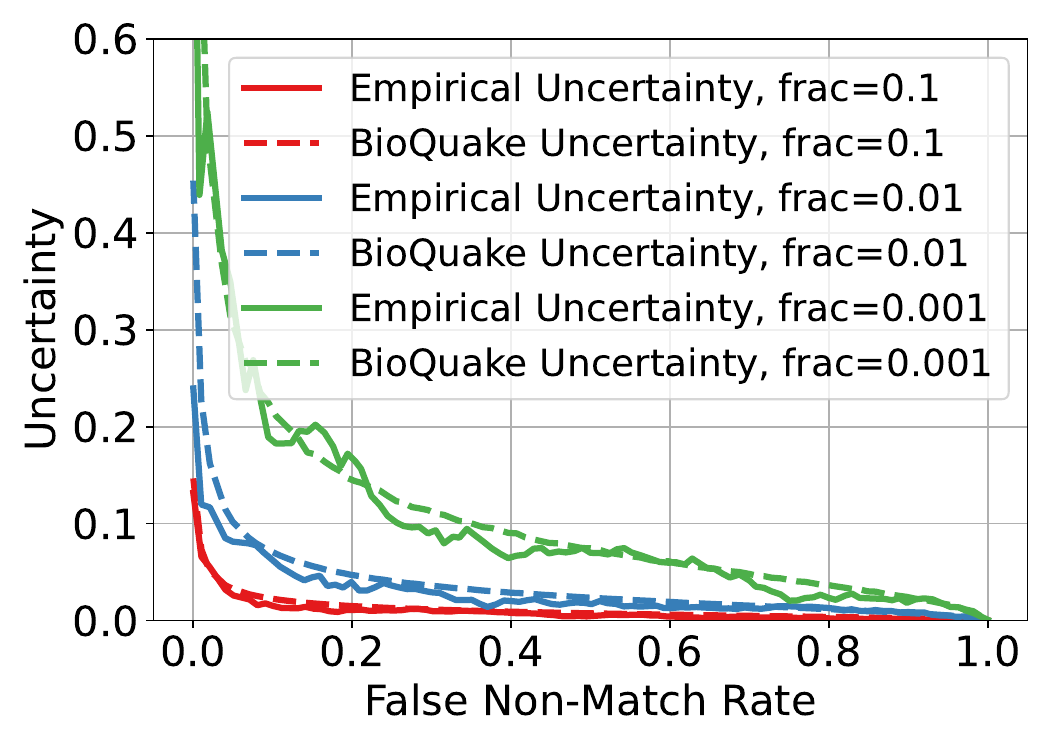}\label{fig:qmagface_adience}} \\
    
    \subfloat[MagFace-ColorFERET]{\includegraphics[width=0.33\textwidth]{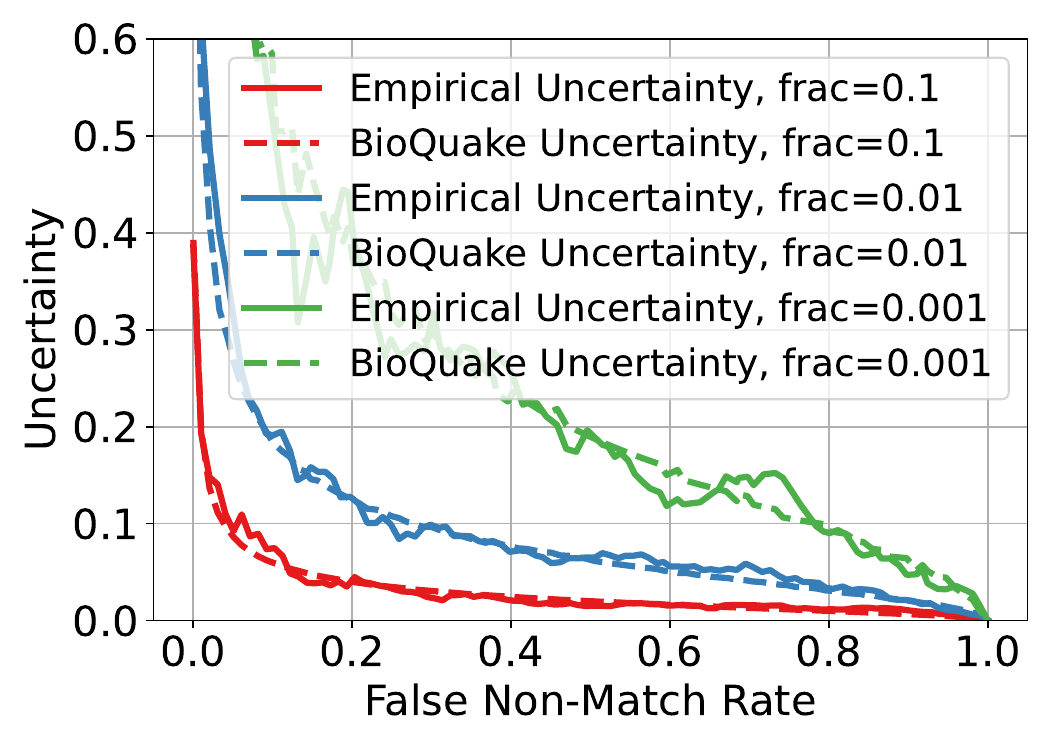}\label{fig:magface_colorferet}} \hfil
    \subfloat[MagFace-LFW]{\includegraphics[width=0.33\textwidth]{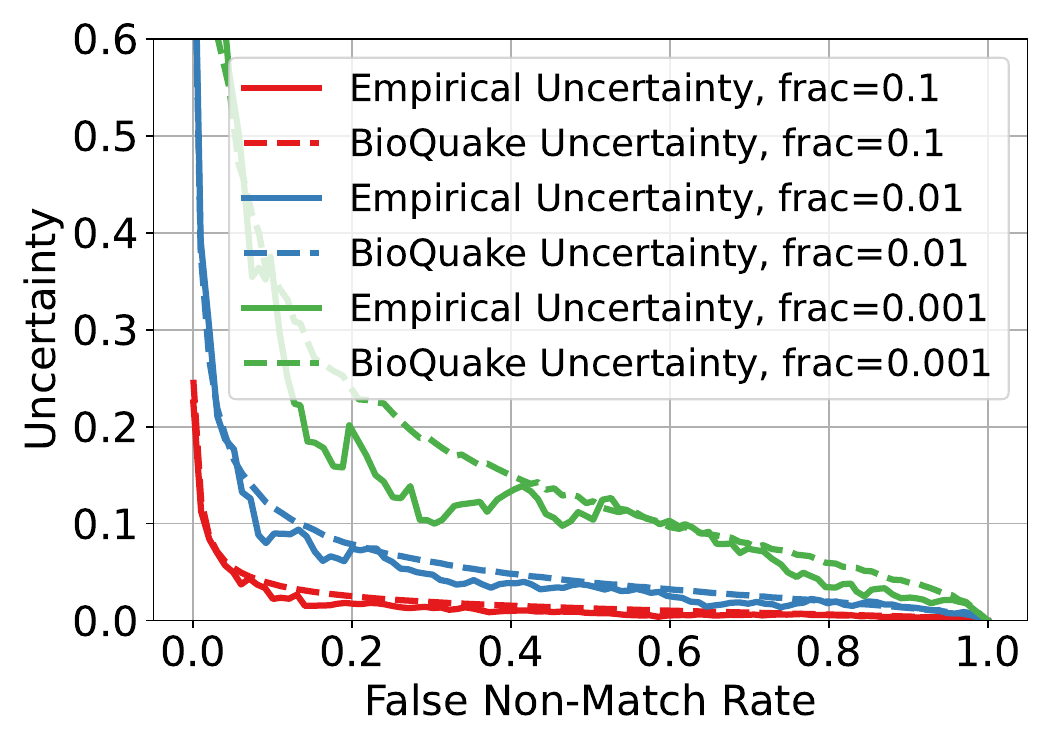}\label{fig:magface_lfw}} \hfil
    \subfloat[MagFace-Adience]{\includegraphics[width=0.33\textwidth]{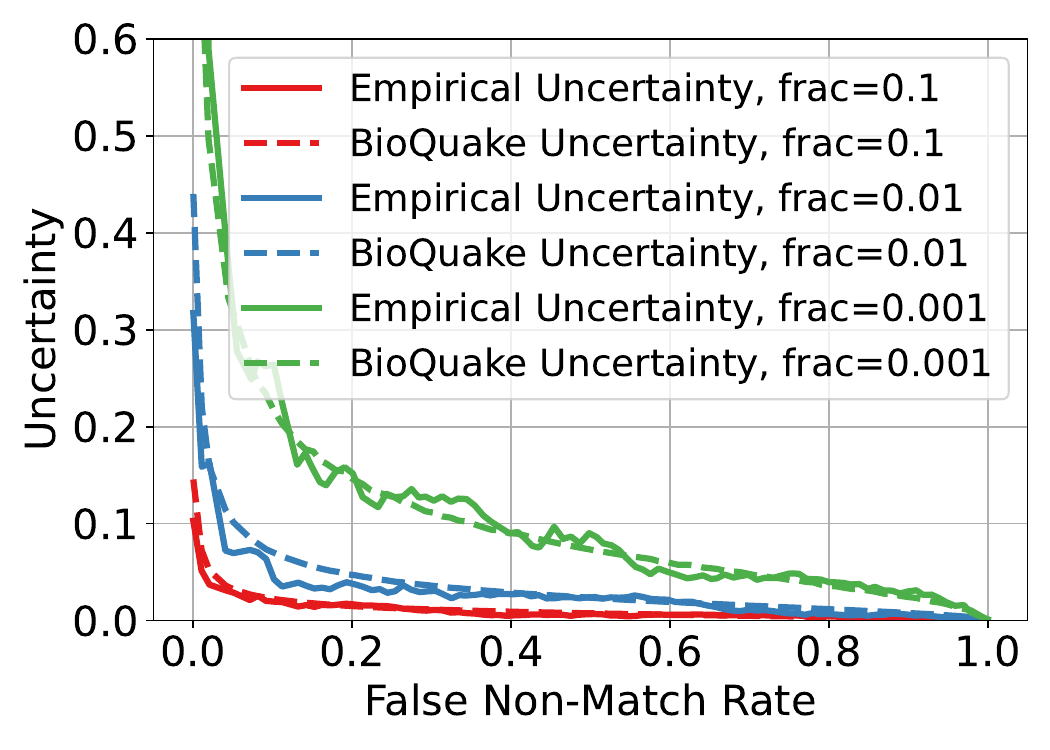}\label{fig:magface_adience}} \\
    
    \subfloat[ArcFace-ColorFERET]{\includegraphics[width=0.33\textwidth]{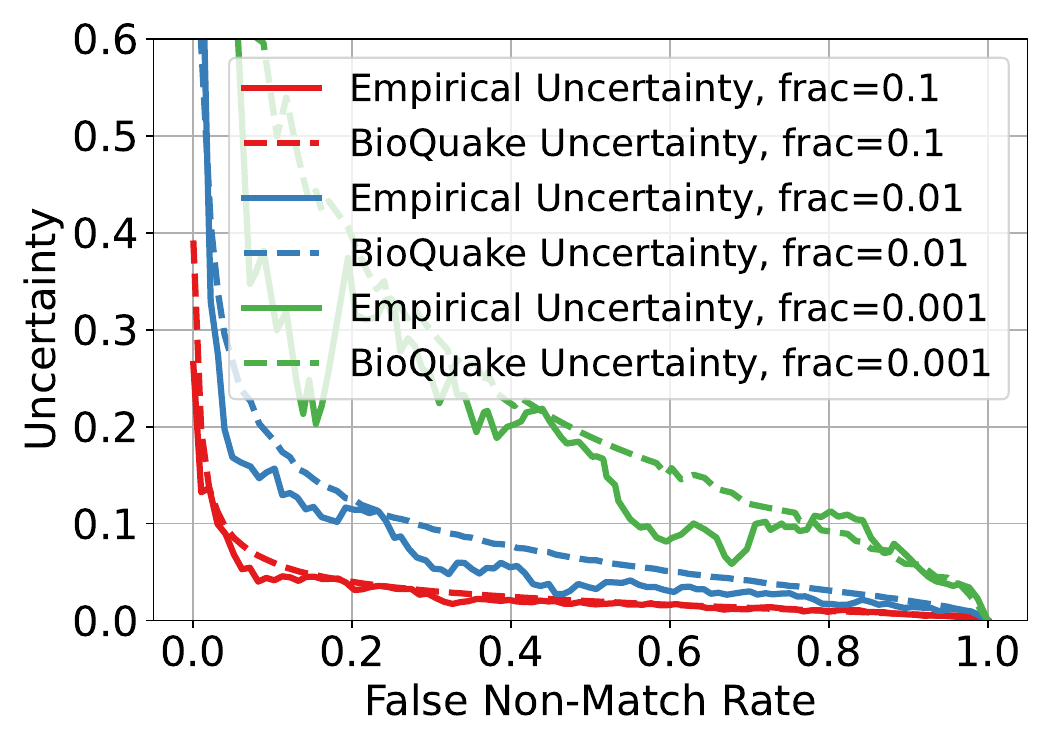}\label{fig:arcface_colorferet}} \hfil
    \subfloat[ArcFace-LFW]{\includegraphics[width=0.33\textwidth]{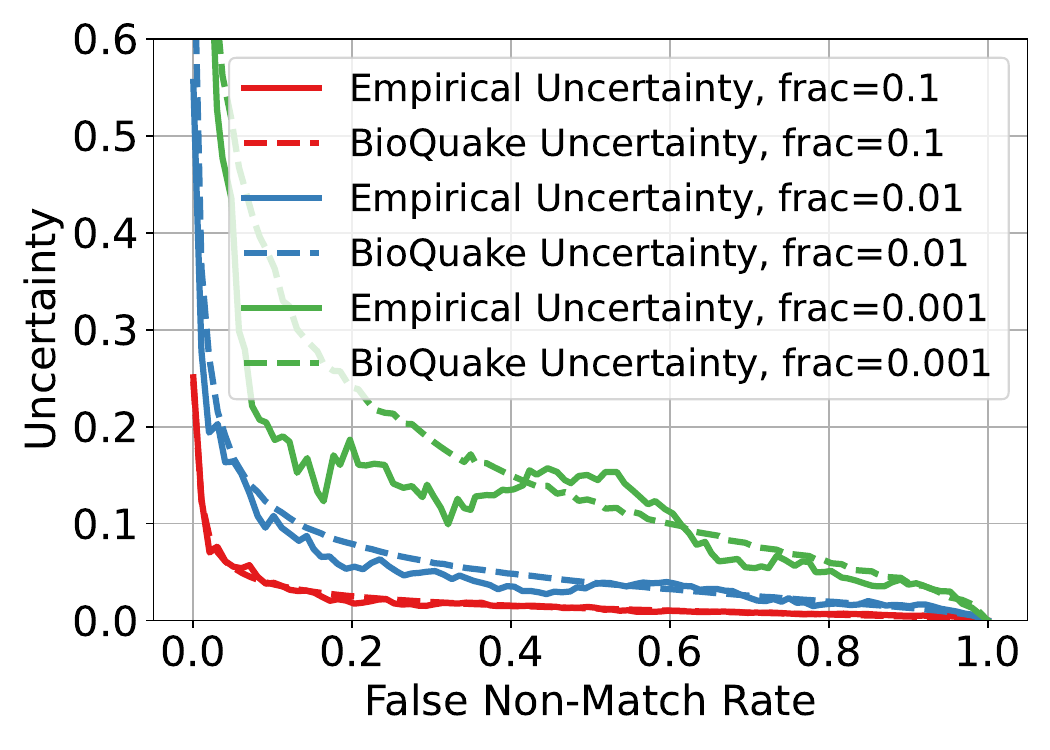}\label{fig:arcface_lfw}} \hfil
    \subfloat[ArcFace-Adience]{\includegraphics[width=0.33\textwidth]{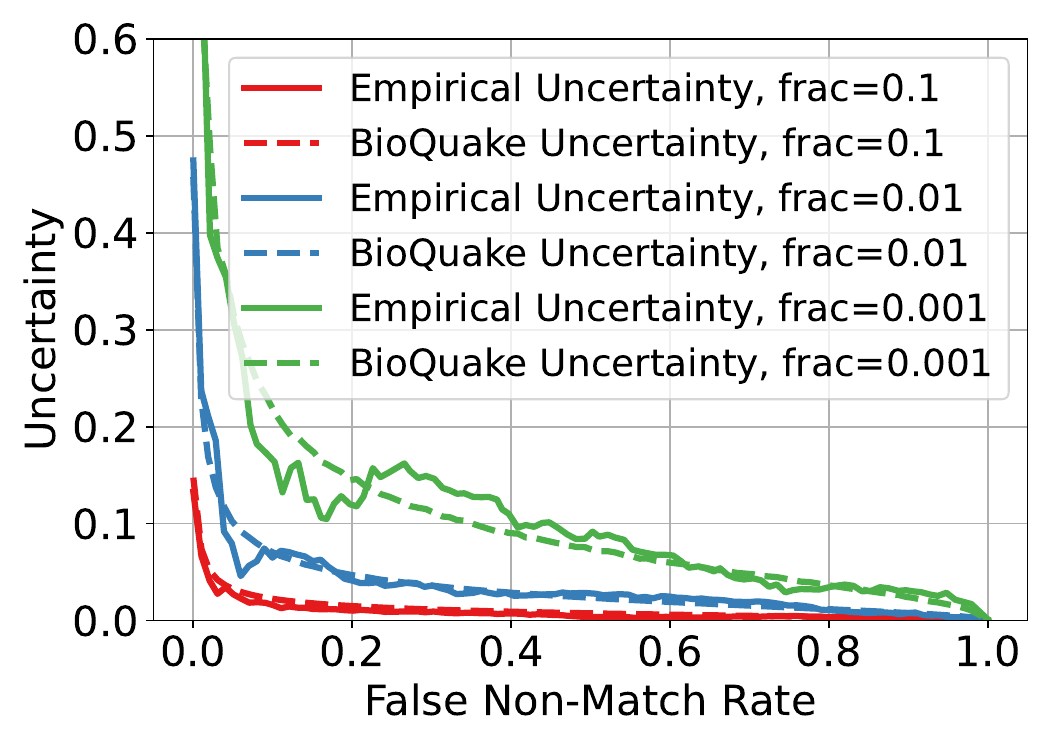}\label{fig:arcface_adience}} \\
    
    \subfloat[FaceNet-ColorFERET]{\includegraphics[width=0.33\textwidth]{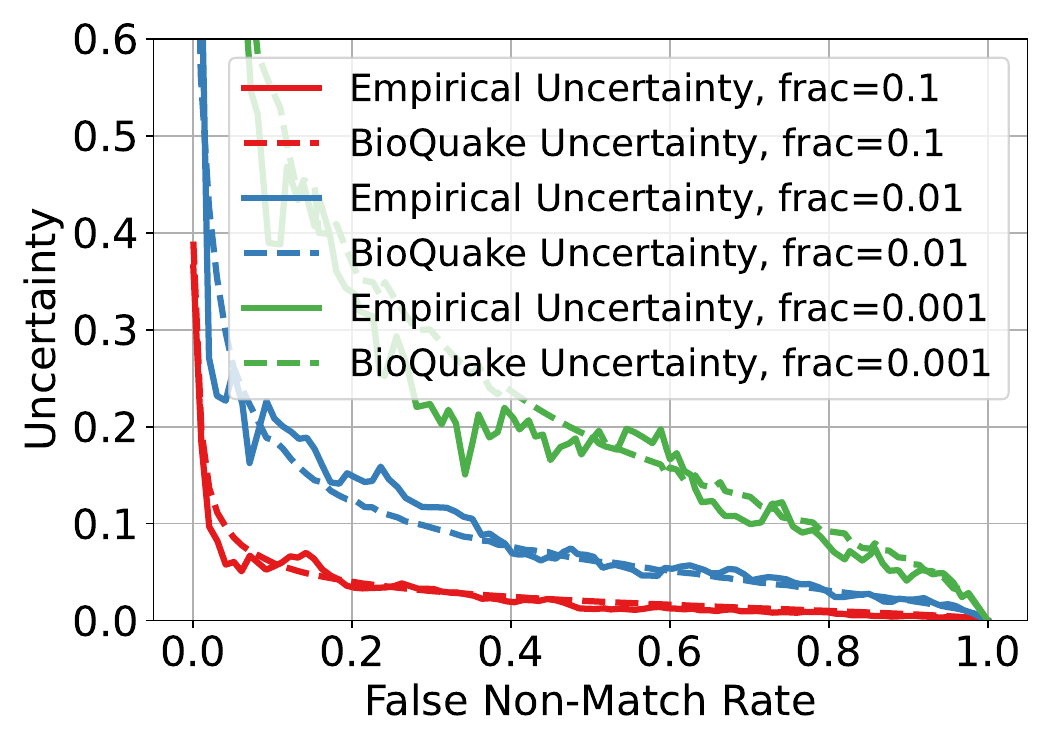}\label{fig:facenet_colorferet}} \hfil
    \subfloat[FaceNet-LFW]{\includegraphics[width=0.33\textwidth]{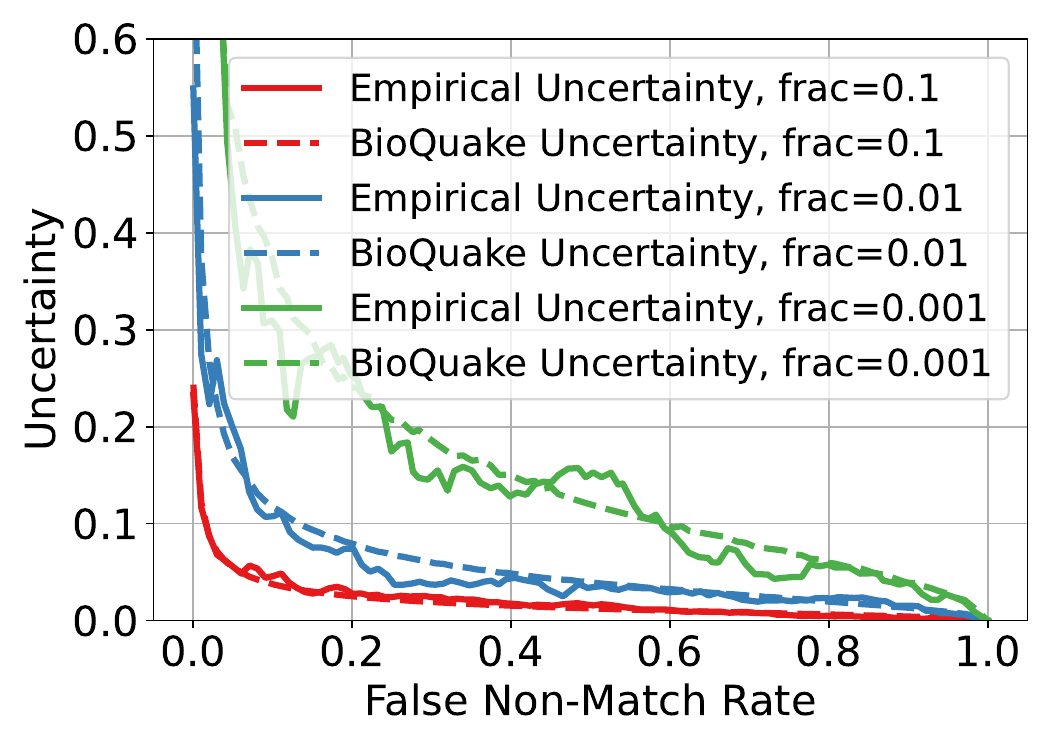}\label{fig:facenet_lfw}} \hfil
    \subfloat[FaceNet-Adience]{\includegraphics[width=0.33\textwidth]{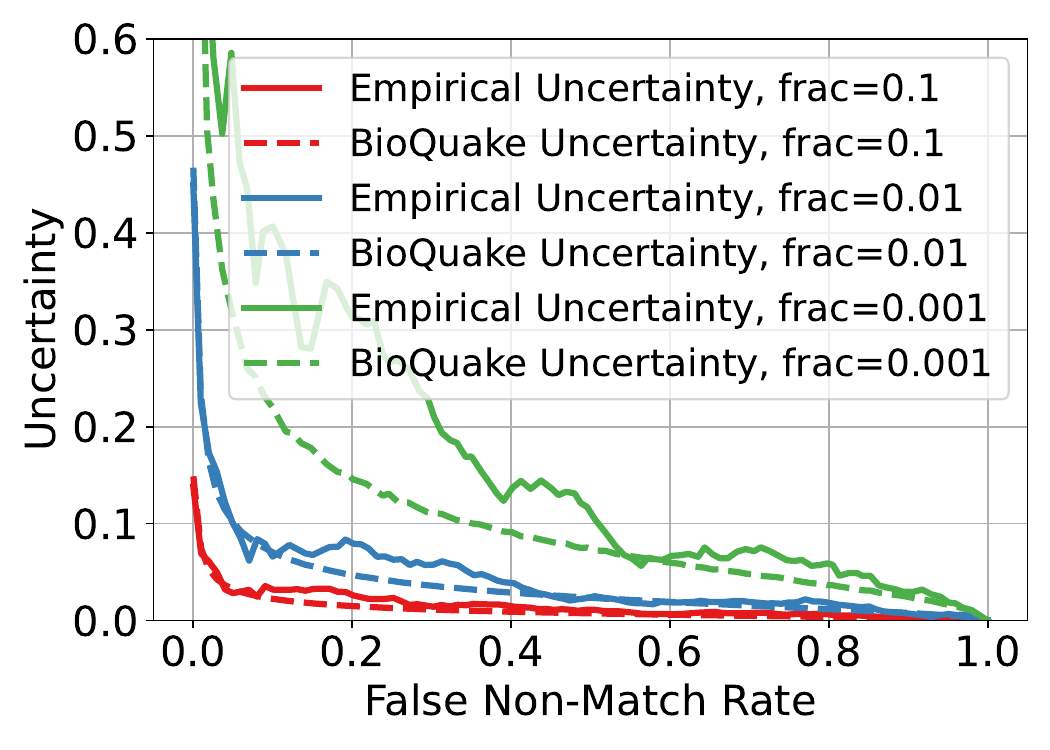}\label{fig:facenet_adience}} 

    \caption{\textbf{Comparative Analysis between the theoretical BioQuake and empircal Uncertainty for FNMRs} - For different fractions (\textit{frac}) of the base datasets, the proposed theoretical BioQuake approach (dashed line) is compared against the empirical uncertainty (solid line) of FNMRs on different model-dataset combinations. A high similarity between both approach is seen indicated the strong effectiveness of BioQuake in estimating the uncertainty.
    }
    \label{fig:fru2}
\end{figure*}

\textbf{Comparative Analysis:} 
To further analyse how well BioQuake can estimate the uncertainty for recognition tasks, BioQuake and the empirical uncertainty are calculated for different fractions of the dataset.
Figures \ref{fig:fru1} and \ref{fig:fru2} show these uncertainties for FMR and FNMR over three datasets and four face recognition models.
In most cases, it can be observed that BioQuake closely resembles the empirical uncertainty.
However, when only a small fraction of the data is used, in some minor cases, such as in Figure \ref{fig:fru2} QMagFace-ColorFeret, MagFace-LFW, ArcFace-ColorFeret, and ArcFace-LFW, the prediction deviates from the experimental values. This can be explained by the stochastic nature of the experimental setup as only a small fraction is used and thus, the data characteristics can vary significantly.
Generally, the results show a high effectiveness of BioQuake in estimating the uncertainty for recognition tasks.

\subsection{Uncertainty Analysis of Existing Datasets}
\label{sec:resultsUncertainty}
To investigate the reliability of existing datasets and benchmarks, we calculate the BioQuake uncertainty on the state-of-the-art verification solutions on various biometric datasets and benchmarks.
Table~\ref{tab:RelatedWork} summarizes this uncertainty analysis.
It involves 62 biometric recognition datasets \textcolor{black}{\footnote{Note that a dataset may be evaluated differently by various recognition systems. In fact, authors might use all possible genuine-impostor pairs or specific subsets of the dataset depending on the evaluation strategy (e.g., reserving certain subsets for training). We consider each configuration as a dataset instance.}}    covering 8 biometric modalities, namely ECG, EEG, eye-tracking, face, fingerprint, gait, iris, and keystroke dynamics. Moreover, it reports the database statistics including the number of captured sessiona (\#S), the number of identities (IDs), and the number of genuine (Gen) and imposter comparisons (Imp) involved in the performance calculation.
The last column in Table~\ref{tab:RelatedWork} shows the minimum reportable errors for datasets according to our 6\% rule. 

{The analysis shows that 24 out of 62 FNMRs and 16 out of 62 FMRs reported on these dataset instances exhibit a \textit{BioQuake} exceeding 0.3, indicating that the error might vary by more than 30\% compared to the reported value, which is the threshold we set for a ``Good'' reliability. More importantly, 18 FNMR (29\%) and 11  FMR (18\%)} instances show a \textit{BioQuake} higher than 0.5, highlighting the need more reliability in biometric verification research.

The last column in Table~\ref{tab:RelatedWork} shows the minimum reportable errors for the datasets when the confidence level is 95\% and the BioQuake is 0.061 (6\% rule). For 15 reported FNMR and 17 reported FMR results, the minimum reportable error exceeds 10\%, indicating significant uncertainty in these measurements. Furthermore, when adopting an FMR of 0.001 as recommended by NIST 800-63B~\cite{grassi2020digital} and the European Border Guard Agency Frontex \cite{FrontexBestPractice} for biometric authentication, only 23 out of the 62 reported errors meet this criterion based on the lowest reportable FMR and a BioQuake of 6.1\%.

In contrast to the other modalities, the datasets for face verification generally demonstrate superior reliability because of their larger size. 
Some exceptions are the LFW-driven benchmarks \cite{huang2008labeled}, which provide a low {number of pairwise (impostor-genuine) comparisons (3000) for both the FNMR and the FMR. }Consequently, no low error rates can be reliably achieved on these small benchmark sizes.
Similarly, some datasets within each modality may not support the prediction of reliable results, even if the performance reported on them suggests otherwise, due to uncertainty tied to the dataset's limitations.

Another aspect highlighted by Table \ref{tab:RelatedWork} is that the same datasets can yield varying comparison scores depending on the evaluation method used. For instance, in the ECG modality, datasets such as ECG-ID \cite{lugovaya2005biometric}, PTB \cite{bousseljot1995nutzung}, CYBHi \cite{da2014check} , and In-house \cite{melzi2023ecg}, as well as the FVC2002 \cite{maio2002fvc2000} in fingerprint analysis, exhibited different comparison scores for the same dataset. 
In \cite{melzi2023ecg}, Melzi et al. consider single and multiple session scenarios, leading to varied comparison scores for each scenario or in \cite{melzi2023ecg}, a dataset containing 25,000 subjects was used but only 5000 genuine pairs were utilized in evaluations.
Consequently, possessing a large dataset alone is insufficient as the number of comparisons utilized is a key factor to consider.

Lastly, we want to point out that many biometric datasets for ECG and EEG, were collected in a single session.
Consequently, these datasets can only be used for testing purposes but are often used with train-test splits.
Future work in this area needs to keep a subject-exclusive train and test data split into account to prevent overfitting as observed in \cite{DBLP:journals/istr/ChaurasiaFSTC24}.

In summary, Table \ref{tab:RelatedWork} shows that many datasets across various biometric modalities are just too small for the high performance current algorithms can achieve, resulting in performance reports of low reliability that are hardly comparable.

\section{Limitations}
\label{sec:limitations}
Despite the effectiveness of BioQuake in practice, there are some limitations to consider. One key assumption is that the samples are treated as independent and identically distributed (iid). While our empirical analysis shows that this assumption does not significantly impact BioQuake's performance in most cases, issues may arise when the number of subjects is small relative to the number of samples, or when the majority of samples are concentrated on only a few subjects. In such scenarios, BioQuake's uncertainty estimation may be affected, as the number of subjects is not directly considered in the estimation process. However, since BioQuake bases its estimation on the number of comparisons rather than the number of subjects, we avoid the need to explicitly model how samples are compared. This simplifies the uncertainty estimation significantly compared to prior methods, while maintaining accuracy.

\section{Conclusion}
\label{sec:Conclusion}
To improve the reliability of biometric verification reporting, we identify two main challenges: first, convincing the research community of the importance of reporting uncertainties, and second, providing an easy-to-use metric for this purpose. This paper introduces and validates BioQuake, a metric for quantifying uncertainty in biometric error rates. Using BioQuake, we analyzed the performance uncertainty of biometric models utilizing 62 different datasets and published in top-tier venues, none of which included reliability information. In up to 29\% of these systems, the reported impostor acceptance rates vary by more than 50\%, underscoring issues with current practices. We conclude with a call to standardize reliability reporting in biometric research to enable more accurate evaluations and faster progress, 
supported by our BioQuake framework as a practical tool for this purpose.

\section*{Acknowledgement}
This work was funded by the Topic Engineering Secure Systems of the Helmholtz Association (HGF) and supported by KASTEL Security Research Labs, Karlsruhe, and Germany’s Excellence Strategy (EXC 2050/1 ‘CeTI’; ID 390696704).
Portions of the research in this paper use the FERET database of facial images collected under the FERET program. 
\bibliographystyle{IEEEtran}
\bibliography{literature}

\end{document}